\documentclass[%
reprint,
superscriptaddress,
amsmath,amssymb,
aps]{revtex4-1}
\usepackage{graphicx}
\usepackage{dcolumn}
\usepackage{bm}
\usepackage{braket}
\usepackage{amsmath}
\usepackage{graphicx}
\usepackage[colorinlistoftodos]{todonotes}
\usepackage{mathtools}
\usepackage{enumerate}

\begin{document}
\title{%
Surface-polaritonic phase singularities and multimode polaritonic frequency combs via dark rogue-wave excitation in hybrid plasmonic waveguide}
\author{Saeid Asgarnezhad-Zorgabad}
 \affiliation{Department of Physics, Sharif University of Technology, Tehran 11365 11155, Iran}
\affiliation{Institute for Quantum Science and Technology, University of Calgary, Calgary, Alberta T2N 1N4, Canada}
\author{Rasoul Sadighi-Bonabi}
\affiliation{Department of Physics, Sharif University of Technology, Tehran 11365 11155, Iran}
\author{Bertrand Kibler}
\affiliation{Laboratoire Interdisciplinaire Carnot de Bourgogne (ICB), UMR 6303 CNRS-Universit\'{e} Bourgogne Franche-Comt\'{e},
21078 Dijon, France}
\author{\c{S}ahin Kaya \"{O}zdemir}
\affiliation{Department of Engineering Science and Mechanics,
The Pennsylvania State University, PA 16802-6812, USA.}
\affiliation{Materials Research Institute, The Pennsylvania State University, University Park, PA 16802-6812, USA.}
\author{Barry C. Sanders}
\email{sandersb@ucalgary.ca}
\homepage{http://iqst.ca/people/peoplepage.php?id=4}
\affiliation{Institute for Quantum Science and Technology, University of Calgary, Calgary, Alberta T2N 1N4, Canada}%
\date{\today}
\begin{abstract}
Material characteristics and input-field specifics limit controllability of nonlinear electromagnetic-field interactions. As these nonlinear interactions could be exploited to create strongly localized bright and dark waves, such as nonlinear surface polaritons, ameliorating this limitation is important. We present our approach to amelioration,
which is based on a surface-polaritonic waveguide reconfiguration that enables excitation, propagation and coherent control of coupled dark rogue waves having orthogonal polarizations.
Our control mechanism is achieved by finely tuning laser-field intensities and their respective detuning at the interface between the atomic medium and the metamaterial layer. In particular, we utilize controllable electromagnetically induced transparency (EIT) windows commensurate with  surface-polaritonic polarization-modulation instability to create symmetric and asymmetric polaritonic frequency combs associated with dark localized waves.
Our method takes advantage of an atomic self-defocusing nonlinearity and dark rogue-wave propagation to obtain a sufficient condition for generating phase singularities. Underpinning this method is our theory which incorporates dissipation and dispersion due to the atomic medium being coupled to nonlinear surface-polaritonic waves~(SPWs). Consequently, our waveguide configuration acts as a bimodal polaritonic frequency-comb generator and high-speed phase rotator,
thereby opening prospects for phase singularities in nanophotonic and quantum communication devices.
\end{abstract}
\maketitle
\section{Introduction}
Controllable excitation for nonlinear plasmonics~\cite{kauranen2012nonlinear} and for polaritonic frequency-comb generation in nanophotonic circuits would be valuable for spectroscopy~\cite{Picqu2019frequency}, quantum~\cite{PhysRevLett.101.130501,tame2013quantum} and fast optical communication~\cite{gaeta2019photonic}. Polaritonic and plasmonic excitation boost nonlinearities due to strong coupling of surface-polaritonic waves~(SPWs) to interface, giant surface-field confinement~\cite{Takahara:97,zayats2005nano}, anomalous spectral responses to the surface optical properties and ultra-fast temporal action of plasmon excitation to the polarization of hybrid interface~\cite{kauranen2012nonlinear}. Recent investigations reveal excitation and propagation of nonlinear surface polaritonic~(plasmonic) waves and explore applications to various nanoplasmonic systems~\cite{yu2019plasmon} such as efficient high-harmonic generation~\cite{chervy2016high,PhysRevLett.120.203903}, ultra-fast dynamics of SPWs~\cite{marini2013ultrafast}, ultra-short pulse focusing~\cite{PhysRevLett.105.116804,Pusch:13},
light spin coupled to plasmon orbit~\cite{PhysRevX.9.021031} and frequency-comb generation~\cite{PhysRevB.98.115410}.

Excitation and propagation of the plasmon oscillation and consequently linear and nonlinear SPWs are limited due to high-Ohmic loss of the metallic layer~\cite{li2016figure}, dissipation of the material layer~\cite{Boriskina:17} and input driving field~\cite{Fisher:15}. In the past few decades, experimental and theoretical investigations~(for an intuitive explanation of dealing with plasmonic loss for waveguide application, see~\cite{khurgin2015deal}) report stable propagation of linear and nonlinear SPWs employing ultra-low loss metallic-type layers such as single-crystal~\cite{Uchino:18} and mono-crystal~\cite{Fedotov:12} metallic film, structured Fano metamaterials~\cite{Cao:12}, semiconductor metamaterials~\cite{naik2011comparative} and superconducting metamaterials~\cite{lazarides2018superconducting}. These investigations reveal that plasmonic excitation and stable propagation need minimal metallic nanostructure roughness~\cite{de2010amplification}. Therefore, polaritonic frequency combs, and generally space-time control of nonlinear surface-polaritonic waves, are unfortunately challenging due to material limitations and driving field characteristics~\cite{PhysRevLett.101.263601,PhysRevA.81.033839,PhysRevA.85.050303}.

Propagation of an optical pulse in nonlinear media leads to the appearance
of strongly localized bright and dark waves such as soliton~\cite{RevModPhys.78.591}, rogue waves and breathers in nearly conservative systems~\cite{akhmediev1997solitons}. Bright rogue waves and breathers are highly localized nonlinear solitary waves with oscillatory amplitudes ~\cite{dudley2014instabilities,dudley2019rogue}. These waves are valuable for their applications to phase and intensity modulation schemes~\cite{PhysRevLett.107.253901,PhysRevX.5.041026}, as well as the formation of bound states and molecule-like behavior~\cite{PhysRevLett.122.084101}. More generally, dissipative rogue waves and breathers~\cite{akhmediev2016roadmap} have potential applications to nonlinear systems such as mode-locked lasers~\cite{ryczkowski2018real} and frequency-comb generators~\cite{yu2017breather}. By contrast, dark rogue waves were only observed during multimode polarized light propagation in a telecommunication fiber to date~\cite{frisquet2016optical,PhysRevA.97.013852}. Besides, their application remains unexplored.

Previous investigations show that the interface between a dielectric and a metallic layer is highly nonlinear and SPWs hence can propagate as a various types of nonlinear optical waves such as soliton, rogue waves and breathers~\cite{PhysRevA.91.023803,PhysRevA.98.013825}. On the other hand, many proposals indicate that the optical properties of the linear and nonlinear SPWs can be controlled~\cite{PhysRevA.81.033839,PhysRevA.85.050303,PhysRevA.91.023803}, and stable propagation of surface polaritonic solitons, rogue waves and breathers can be achieved by employing a hybrid plasmonic waveguide comprising a negative index-metamaterial~(NIMM) layer
and a thin atomic medium layer~\cite{PhysRevA.98.013825}.

Excited bright surface-polaritonic breathers have applications to plasmonic-phase modulation~\cite{melikyan2014high} and polaritonic frequency-comb generation~\cite{PhysRevA.99.051802,geng2016frequency}. Therefore, natural questions that appear are whether the dark-surface polaritonic rogue waves can be excited by stable propagation and nonlinear interaction of multimode SPWs, whether they are controllable and what would be the application of these nonlinear polaritonic dark rogue waves. The existence of coupled dark rogue waves, their coherent control and resultant applications to an experimentally feasible hybrid waveguide has not yet been investigated.  

We ameliorate the controllability limitation arising due to material characteristics and input-field properties by proposing a hybrid plasmonic waveguide that exploits spectral control of nonlinear electromagnetic-field interactions including surface polaritons. Our waveguide enables excitation, propagation and coherent control of coupled dark rogue waves with orthogonal polarization and includes finely tuned laser field intensities and correspond detuning at the interface between atomic medium and NIMM layer. We exploit controllable double electromagnetically induced transparency (DEIT) windows~\cite{wang2017strong} commensurate with plasmonic analogue of polarization-modulation instability to create symmetric surface polaritonic frequency combs associated with dark rogue waves. Furthermore we take advantage of atomic self-defocusing nonlinearity at EIT windows to obtain the sufficient condition for surface-polaritonic phase singularities.

Consequently, our waveguides twist surface polaritonic phase and generate controllable bimodal frequency combs based on generating phase singularities and coupled dark rogue wave excitations, thereby opening prospects for designing ultra-fast phase rotor and multimode frequency-comb generator for nanophotonic and quantum optical communication devices.
Our method for generating, controlling and propagating polaritonic dark rogue waves, surface polaritonic phase singularities and multimode polaritonic frequency combs is based on introducing atomic dissipation and dispersion to coupled nonlinear SPWs and is novel.

The rest of our paper is organized as follows. In \S\ref{background} we present the background of our works. The stable excitation of coupled SPs and linear propagation regime is expressed in \S\ref{linearExcitation} and we explore the coupled nonlinear SPP propagation and dark rogue-wave formation in \S\ref{nonlinearSPP}. Finally, we discuss and summarize our results in \S\ref{Discussion} and \S\ref{conclusions}, respectively.

\section{\label{background}Background}
We begin by briefly reviewing DEIT windows.
Next, we discuss polarization-modulation instabilities and
introduce the Manakov system and bright and dark rogue-wave formation. Finally, we review salient aspects of generation and propagation frequency combs.

\subsection{\label{DEIT_windows}Double electromagnetically induced transparency}
This subsections start with pertinent basic concepts of the single-~\cite{peng2014and,PhysRevLett.107.163604}- and double~\cite{PhysRevA.89.021802,wang2017strong,PhysRevA.94.053832}-induced transparency windows. We discuss the important properties of the spectral transparency windows necessary for the generation of self-focusing/self-defocusing and cross-focusing/cross-defocusing nonlinearities. Self-defocusing and cross-defocusing nonlinearities are necessary to generate coupled-dark rogue waves and phase singularities.

The notion of an electromagnetically induced transparency window refers to interfering electronic transition of an atomic medium to eliminate or reduce resonant atomic absorption and modulate the linear dispersion. This quantum interference is a result of Fano interference~\cite{PhysRev.124.1866} that requires coupling of discrete transition to a continuum and creates narrow spectral transparency windows. Dispersion of EIT-assisted atomic medium with detuning~($\delta$), eigenfrequencies~($\delta_{\pm}$) and dispersion constants~($\chi_{\pm}$)
\begin{equation}
    \chi=\frac{\chi_{+}}{\delta-\delta_{+}}+\frac{\chi_{-}}{\delta-\delta_{-}},
\end{equation}
represents strong Fano interference and hence strong absorption reduction in the resonant condition with zero detuning~($\delta=0$).

Consequently, DEIT windows extend the concept of electromagnetically induced transparency to create two transparency windows for a probe field over different narrow spectral domains. Controlling spectral widths of DEIT windows is important for applications to coherent control of light amplification, enhancing nonlinear optical susceptibilities, to achieve coherent control of frequency widths for transparency windows in order to create strong self- and cross-defocusing optical nonlinearities (i.e., self- and cross-phase modulation) and to manipulate group-velocity dispersion of the atomic medium. 

\subsection{\label{Polarization_modulation_instability} Modulation instability and dark rogue waves}
This subsection begins by introducing the concept of the modulation instability~\cite{zakharov2009modulation}. We then discuss the properties of modulation instability in homogeneous media with normal/anomalous dispersion. Finally, we briefly explain the polarization-modulation instability as the possible origin of the coupled nonlinear waves such as dark rogue waves. Our discussion in this subsection ends by introducing the Manakov system, which describes dynamics of the coupled nonlinear dark waves. 

Modulation instability is known as a process in which a weak periodic perturbation can be amplified through propagation in a nonlinear medium~\cite{zakharov2009modulation}. In the scalar description of electromagnetic waves or single-mode propagation regime, modulation instability occurs only for anomalous dispersion and the process is described by the common nonlinear Schr\"odinger equation (NLSE). However, for a vector electromagnetic field, which can be described by a pair of two-circularly polarized electromagnetic fields within the medium, modulation instability can emerge in both normal and anomalous dispersion regimes. This nonlinear process is then usually termed polarization modulation instability~\cite{Millot:14}. The first multicomponent NLSE type of model with applications to physics is the well known Manakov model ~\cite{manakov1974theory}. For two polarization components, the corresponding set of two coupled NLSEs is completely integrable. In this framework~\cite{PhysRevA.92.053854}, polarization-modulation instability was found to be the origin of coupled bright or dark rogue waves~\cite{PhysRevA.97.013852,chen2017versatile}. 

Nonlinear dynamics of the two-mode electric field with amplitudes $q_{1,2}$ and formation of rogue waves are described by the following system of coupled equations~\cite{manakov1974theory}
\begin{align}
    \text{i}\frac{\partial q_n}{\partial x}+\frac{1}{2}\frac{\partial^2q_n}{\partial \tau^2}+\left(|q_1|^2+|q_2|^2\right)q_n=0, \; n\in\{1,2\}.
    \label{Main_Manakov}
\end{align}
This coupled system can be  extended to describe other physical systems by adding higher-order effects to coupled nonlinear Schr\"odinger equations~\cite{du2018rogue}.
However, we neglect effects due to these nonlinear dynamical evolution terms in the coupled nonlinear SPWs as these nonlinear terms are related to higher-order dispersion and nonlinearities, thereby leading to distinct rogue-wave and polarization-modulation instability properties. Moreover, the rogue waves can be generated and propagated in the coupled nonlinear Schr\"odinger equation with nonlinear coherent coupling term~\cite{zhang2018rogue}. The positive nonlinear coherent coupling term leads to energy exchange between the two propagating modes whose effects on coupled nonlinear SPWs goes beyond the scope of this work. In our work, we employ the standard Manakov system~(\ref{Main_Manakov}) to investigate dynamical evolution of the coupled nonlinear SPWs through the interface between the atomic medium and the metamaterial layer.

\subsection{\label{Frequency_combs}Frequency combs}
This subsection introduces key concepts of optical frequency combs. Frequency combs refers to equidistant optical frequency components associated with a regular train of ultrashort pulses (with high degree of mutual coherence) at a fixed repetition rate, and may be generated in an optical resonator especially~\cite{RevModPhys.75.325,PhysRevLett.100.013902,del2007optical,kippenberg2011microresonator}, but not exclusively. In an optical microresonator with off-set frequency $f_\text{O}$ and pulse repetition rate $f_\text{r}$, the $n$th frequency of the optical modes is
\begin{equation}
    f_n=n\times f_\text{r}+f_\text{O}.
\end{equation}
Coherent frequency comb excitation would be valuable due to their wide applications to quantum optics and information~\cite{kues2019quantum}, spectroscopy~\cite{picque2019frequency}, optical clock~\cite{papp2014microresonator} inter alia.

Single-mode surface polaritonic frequency combs can be efficiently excited in a nonlinear medium with reduced group-velocity dispersion~\cite{PhysRevA.99.051802}. The frequency combs in a weakly perturbed wave within a nonlinear medium with higher-order dispersion $\beta_n(\omega)$
and frequency~$\delta$ are
obtained by~\cite{gaeta2019photonic}
\begin{equation}
    \omega_\delta
    	=\omega_\text{O}+\sum_n\frac{D_n}{n!}\delta^n,\;
    D(\delta)=\sum_{n=2,3,\cdots}\frac{\beta_n}{n!}\delta^n,
\end{equation}
which represents nonlinear dispersion. We establish controllable excitation and propagation of two-mode surface polaritonic frequency combs,
and we demonstrate formation of coupled-dark rogue wave by symmetric frequency-combs generation within DEIT windows.    

\section{Approach}
We first qualitatively
model our plasmonic waveguide in~\S\ref{model} by introducing the NIMM layer, the atomic medium and driving fields.
Second, in~\S\ref{Mathematics},
we introduce an analytical quantitative mathematical description of our waveguide by employing a macroscopic description to NIMM layer, treating driving lasers as semi-classical fields and obtaining SPP dynamics using the Maxwell-Bloch equation. Finally, we describe our perturbative approach and transformation to solve Maxwell-Bloch equations approximately
in~\S\ref{Methods}.
\subsection{\label{model}Model}
In this subsection, first we introduce our waveguide configuration. Next we discuss the NIMM layer, atomic medium and briefly discuss the possible realistic models of these materials for our proposed scheme. Finally, we explain the irradiation and coupling of laser fields and microwave fields to the interface between the atomic medium and the metamaterial.   

\begin{figure}
\includegraphics[width=\columnwidth]{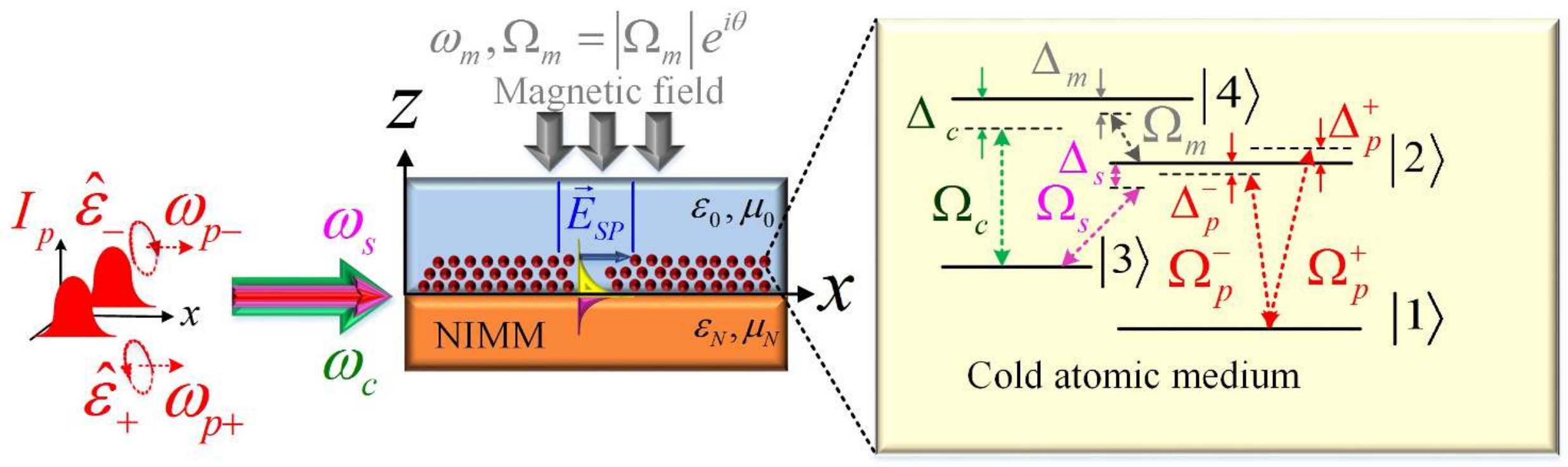}
\caption{\label{fig:one}
Proposed multimode nonlinear waveguide, comprising a $N$-type atoms doped into a lossless dielectric placed above a loss-free NIMM layer. Copropagating coupling~(c), signal~(s) and orthogonally-polarized weak probe~(p) lasers and a $\mu$-wave ($\mu$) field drive the system with Rabi frequencies $\Omega_\text{c}$~(green arrows), $\Omega_\text{s}$~(magenta arrows) and $\Omega_\text{p}^\pm$~(red arrows), and $\Omega_\mu$ (gray arrows), respectively. Detunings from atomic transitions are $\Delta_\text{c,s}$, $\Delta_\text{p}^\pm$, and $\Delta_\text{m}$.}
\end{figure}

Excitation and stable propagation of coupled dark rogue waves are obtained by a nonlinear waveguide as depicted in Fig.~\ref{fig:one}. This polaritonic apparatus consists of two parts. We assume the upper layer as a transparent medium, a NIMM layer as a bottom layer and a coherently driven four-level $N$-type atomic medium~\cite{PhysRevA.84.053820} introduced as a dopant along the dielectric-metamaterial interface. Our configuration serves as a nonlinear waveguide formed by dopant atoms in a lossless dielectric medium over a thickness of several dipole-transition wavelengths. 

Various metamaterial layers such as active and passive NIMM and epitaxial silver films can serve as the plasmonic waveguide in our proposed scheme~\cite{shalaev2007optical,Baburin:19}. Here we assume a fishnet structure with nanorods possessing low-loss in the optical frequency region.
We assume that this NIMM layer is infiltrated with a dipolar gain medium such as a dye and pumped by an additional trigger laser either perpendicularly through the bottom layer or via end-fire coupling technique~\cite{xiao2010loss}.

We choose an N-type atomic medium because of its controllable dispersion and giant Kerr nonlinearity~\cite{PhysRevA.84.053820}. Specifically, we study $\text{Pr}^{3+}$-impurities within a $\text{Y}_2\text{Si}\text{O}_5$ crystal with energy levels~\cite{PhysRevLett.84.4080}
\begin{align}
\ket1=\ket{^3\text{H}_4,F=\pm5/2},\; \ket2=\ket{^1\text{D}_2,F=\pm1/2},\nonumber\\
\ket3=\ket{^3\text{H}_4,F=\pm3/2},\; \ket4=\ket{^1\text{D}_2,F=\pm5/2}.
\label{Eq:one}
\end{align}
Atomic density is~$N_\text{a}$, and natural and dephasing decay rates for $\ket{m}\leftrightarrow\ket{n}$ are~$\Gamma_{mn}$ and~$\gamma_{mn}^\text{dep}$, respectively~\cite{boyd2003nonlinear}. This sample is then cooled with a cryostat to near the liquid-Helium temperature
(GHz regime)
and assumed as a top layer of our proposed hybrid waveguide.

Inhomogeneous broadening, which is negligible in cooled gases due to a weak Doppler effect, includes spin-spin and dipole-dipole interactions. These two interactions are quite large for our solid-state system near the liquid-Helium temperature~\cite{PhysRevA.66.063802}, but inclusion of these interactions in nonlinear SP dynamics requires further consideration, which is beyond the scope of our work. Multimode polaritonic frequency combs and SP phase singularities involve preparing a special ensemble of $\text{Pr}^{3+}$ using persistent spectral hole burning~\cite{ham1997enhanced},
which can circumvent limitations due to inhomogeneous broadening.

Our hybrid waveguide is irradiated by three lasers: a coupling~(c), a signal~(s) and a circularly polarized probe~(p) laser. These laser beams are injected into the interface between the atomic medium and the NIMM layer. We assume that these fields are transversely confined to the interface with a coupling function $\zeta_\text{m}(z)$, $\text{m}\in\{\text{c},\text{s},\text{p}\}$, that drive atomic transitions according to Rabi frequencies~$\Omega_\text{m}$. We assume that these laser fields with different frequencies are produced from a single tunable dye laser and slightly frequency shifted using acousto-optic modulators. Moreover, a microwave (denoted~$\mu$) field with central frequency $\omega_\mu$ is incident perpendicularly to the atomic medium-NIMM layer and drive the hyper-fine atomic transition.

\subsection{\label{Mathematics}Mathematical formalism of the plasmonic waveguide}
This subsection starts with the quantitative description of our waveguide. First, we model our NIMM layer and give a mathematical formalism to describe the NIMM layer. Next we discuss the quantitative description of our atomic medium at the interface and employ the Liouville formalism commensurate with system Hamiltonian to obtain dynamics of the atomic medium interacting with evanescenct laser fields. Finally, we mathematically describe dynamics of the SPWs in our polaritonic waveguide using reduced Maxwell equation coupled to Liouville equation. 

 We evaluate the optical properties of this NIMM layer employing macroscopic description of the metamaterial structure and we describe the permittivity and permeability of this structure using the Drude-Lorentz model~\cite{PhysRevLett.101.263601,xiao2009yellow,sang2017characterization}
with permittivity
\begin{equation}
    \varepsilon_\text{N}=\varepsilon_{\infty}-\frac{\omega_\text{e}^2}{\omega_l(\omega_l+\text{i}\gamma_\text{e})}
\end{equation}
and permeability
\begin{equation}
    \mu_\text{N}=\mu_{\infty}-\frac{\omega_\text{m}^2}{\omega_l(\omega_l+\text{i}\gamma_\text{m})},
    \label{Eq:Optical properties NIMM}
\end{equation}
for~$\varepsilon_{\infty}$
and~$\mu_{\infty}$ the background constant for the permittivity
and permeability,
respectively.
The other constants are $\omega_l$ the perturbation frequency, $\omega_\text{e}$
and~$\omega_\text{m}$ are the electric and magnetic plasma frequencies,
and $\gamma_\text{e}$
and~$\gamma_\text{m}$ are the corresponding decay rates.
We assume that this NIMM layer is filled with a $\text{R}6\text{G}$ dye molecule with macroscopic gain.
The effect of dye-molecule gain 
is incorporated by exploiting the dipole approximation and taking into account the radiative and nonradiative decay rates of the $\text{R}6\text{G}$ dye material~\cite{ford1984electromagnetic}.

We denote atomic energy levels by
\begin{equation}
    E_j=\hbar\omega_j,\;
    j\in\{1,2,3,4\}
\end{equation}
for each atomic level~$\ket{j}$.
Transitions are achieved by driving with three co-propagating laser fields at frequencies $\omega_\text{s,c,p}$ in which~c and~s lasers
drive transitions $\ket3\leftrightarrow\ket4$,
$\ket3\leftrightarrow\ket2$ and orthogonally polarized~p lasers drive $\ket1\leftrightarrow\ket2$ transition,
respectively. These laser fields are injected into our hybrid waveguide employing the end-fire coupling technique~\cite{Maier2005}.
Moreover, a weak $\mu$-wave field with initial phase $\theta$ that is incident perpendicularly to the atomic medium-NIMM interface drives $\ket2\leftrightarrow\ket4$ hyperfine transition.
The corresponding frequency detunings are
\begin{align}
    \Delta_\text{c}:=&\omega_{42}-\omega_\text{c},\;
    \Delta_\text{s}:=\omega_{23}-\omega_\text{s},\nonumber\\
    \Delta_\mu:=&\omega_{42}-\omega_\mu,\; 
    \Delta_\text{p}^{\pm}:=\omega_{42}-\omega_\text{p}^{\pm},
\end{align}
for $\omega^+$ ($\omega^-$) denoting left-~(right)- circularly polarized probe field frequencies.

The electric field of the waveguide is
\begin{equation}
    \bm{E}(\bm{r},t)=\sum_{m=\textrm{c,s,p}}\bm{E}_m(\bm{r},t)+\text{c.c.},
\end{equation}
with
\begin{equation}
    \bm{E}_l(\bm{r},t)=\mathcal{E}_l\bm{u}_l(\bm{r})\exp\{\text{i}(\bm{k}\cdot\bm{r}-\omega_lt)\},
\end{equation}
the electric field of the pumped lasers,
\begin{equation}
    \bm{u}_l(\bm{r})=c\left[k(\omega_l)\bm{e}_{z}-\text{i}k_\text{N}(\omega_l)\bm{e}_{\parallel}\right]/\varepsilon_0\omega_l,
\end{equation}
the electric field vector along the interface,
\begin{equation}
    k(\omega_l)=\frac{\omega_l}{c}\sqrt{\frac{\varepsilon_\text{N}\varepsilon_0(\varepsilon_\text{N}\mu_0-\varepsilon_0\mu_\text{N})}{\varepsilon_\text{N}^2-\varepsilon_0^2}},
\end{equation}
the propagation constant of the SPWs,
\begin{equation}
    \mathcal{E}_l=\left(\frac{\hbar\omega_l}{\varepsilon_0L_{x}L_{y}L_{z}}\right)
\end{equation}
the amplitude of the electric field, $L_{x}$ ($L_{y}$) the length of the atomic medium-NIMM interface, 
\begin{equation}
    L_{z}=\sum_{j=\text{N},0}\left\{\left(\frac{\omega_l^2}{2\text{c}^2}\left[\frac{\tilde{\varepsilon}_j(|\bm{k}_j|^2+|\bm{k}|^2)}{|\bm{k}_j||\varepsilon_j^2|}\right]+\frac{\tilde{\mu}}{2|\bm{k}_j|}\right)\right\}
\end{equation}
the confined mode effective length that determines the confinement of the EM waves to the interface and
\begin{align}
    \tilde{\varepsilon}_j:=&\text{Re}\left[\frac{\partial(\omega_l\varepsilon_j)}{\partial \omega_l}\right], \; \tilde{\mu}_j:=\text{Re}\left[\frac{\partial(\omega_l\mu_j)}{\partial \omega_l}\right],
\end{align}
the effective electrical permeability and magnetic permeability of the interface. 

The Rabi frequencies of the laser fields are
\begin{align}
    \Omega_\text{c}=|\bm{p}_{34}|\mathcal{E}_{c}/\hbar,\;\Omega_\text{s}=|\bm{p}_{23}|\mathcal{E}_{s}/\hbar,\;\Omega_\text{p}^{\pm}=|\bm{p}_{34}|\mathcal{E}_\text{p}^{\pm}/\hbar,
\end{align}
with $\bm{p}_{ij}=|p_{ij}|\bm{e}_{ij}$, where $\bm{e}_{ij}$ denotes the unit vector of the electrical dipole moment, is the dipole moment of the $\ket{i}\leftrightarrow\ket{j}$ atomic transition. These laser fields tightly confined to the interface both transversely and longitudinally according to the evanescent decay function~\cite{PhysRevA.91.023803}
\begin{equation}
    \zeta_\text{p}^{\pm}\approx\zeta_\text{c}\approx\zeta_\text{s}:=\bm{e}_{12}\cdot \bm{u}_\text{p}^{\pm}(\bm{r}),
\end{equation}
and the Rabi frequency of the $\mu$-wave field as
\begin{align}
   \Omega_\mu=|\Omega_\mu|\exp\{\text{i}\theta\}.
\end{align}
deplete the excited state's energy levels and interact with the N-type atomic medium in our waveguide. 

The Hamiltonian of this system in the interaction picture, under rotating-wave and dipole approximations is
\begin{align}
    H_\text{I}=&\hbar\bigg[\sum_{l=1}^{4}\Delta_l\ket{l}\bra{l}+\zeta_\text{c}(z)\Omega_\text{c}\ket{4}\bra{3}\nonumber\\&+\zeta_\text{s}(z)\ket2\bra{3}+\zeta_\text{p}(z)\Omega_\text{p}\ket2\bra{1}+\Omega_\text{m}\ket{4}\bra{2}\bigg]
\end{align}
for $\Delta_1\equiv0$.
The other rotated detunings are
\begin{equation}
    \Delta_2=\Delta_\text{p},\;
    \Delta_3=\Delta_\text{p}-\Delta_\text{s},\;
    \Delta_4=\Delta_\text{c}+\Delta_\text{p}-\Delta_\text{s},
\end{equation}
and we assume $\Delta_\text{m}=\Delta_\text{c}-\Delta_\text{s}$. Dynamics of this hybrid interface is described by the Liouville equation
\begin{equation}
    \text{i}\hbar\left(\frac{\partial}{\partial t}+\Gamma\right)\tilde{\rho}=[H_\text{I},\tilde{\rho}],
    \label{Eq:Liouville Equation}
\end{equation}
for~$\Gamma$ the $4\times4$ matrix describing the decay rates of the $\text{Pr}$:$\text{Y}_2\text{Si}\text{O}_5$ crystal
due to dephasing or other inhomogeneous mechanisms. The explicit solution to the Liouville equation is expressed in Appendix~\ref{Liouville}.

Excitation of surface-polaritonic waves is obtained by tight confinement of the plasmonic field through the interface and modulation of the dissipation and dispersion of the atomic medium. We assume that strong coupling of the weak probe field
to $\ket1\leftrightarrow\ket2$ transition yields exciting and propagating the surface-polaritonic waves
in which dynamics is governed by the Maxwell equation
\begin{equation}
    \nabla^2\bm{E}_\text{p}-\left(\frac{1}{\text{c}^2}\right)\frac{\partial^2\bm{E}_\text{P}}{\partial t^2}=\frac{1}{\varepsilon_0\text{c}^2}\frac{\partial^2\bm{P}_\text{p}}{\partial t^2},
    \label{Eq:Maxwell Equation}
\end{equation}
here the electrical dipole moment of the system is
\begin{equation}
    \bm{P}_\text{p}=N_\text{a}\bm{p}_{21}\tilde{\rho}_{21}\exp\{\text{i}(\bm{k}_\text{p}\cdot\bm{r}-\omega t)\}.
\end{equation}
We employ the slowly varying amplitude approximation as
\begin{equation}
    \left|\frac{\partial^2\Omega_\text{p}}{\partial x^2}\right|\ll\left|2\text{i}k_\text{p}\frac{\partial\Omega_\text{p}}{\partial x}\right|,
\end{equation}
to reduce Eq.~(\ref{Eq:Maxwell Equation}) to
\begin{equation}
    \text{i}\left(\frac{\partial}{\partial x}+\frac{1}{\text{n}_\text{eff}\text{c}}\frac{\partial}{\partial t}\right)\Omega_\text{p}+\kappa_{21}\braket{\tilde{\rho}_{21}}=0,
    \label{Eq:Reduced Maxwell}
\end{equation}
with
\begin{equation}
  \kappa_{21}
    =\frac{N_\text{a}\omega_\text{p}|\bm{e}_\text{p}.\text{p}_{21}|^2}{2\hbar\varepsilon_0\text{c}},
\end{equation}
$\bm{e}_\text{p}$ the unit vector of the polarized electric field, $\text{c}$ the speed of light in free space, $n_\text{eff}=\text{c}k_\text{p}/\omega_\text{p}$ the effective refractive index of the waveguide and
\begin{equation}
    \braket{\psi(z)}=\frac{\int_{-\infty}^{+\infty}\text{d}z\;\zeta^*(z)\psi(z)}{\int_{-\infty}^{+\infty}\text{d}z\;|\psi(z)|^2}.
\end{equation}

We employ the circularly polarized probe laser beam to excite and stably propagate of the multiple mode SPWs. To this aim, we employ 
\begin{equation}
    \bm{e}_{\parallel}\mapsto \hat{\epsilon}_{\pm}
\end{equation}
for right
($\epsilon_+$) and left
($\epsilon_-$) circular polarizations, respectively, with $\epsilon_{\pm}=(\bm{e}_{x'}\pm\text{i}\bm{e}_{y'})/\sqrt{2}$ as unit electric vector of the polarized laser light. Unit vectors possess small rotation to achieve equally confined plasmonic field for each polarization component and we also assume small frequency shift induced by acousto-optic modulators.
We then evaluate dynamics of the Rabi frequency for each polarization components using reduced Maxwell equation~(\ref{Eq:Reduced Maxwell}) commensurate with the Liouville equation~(\ref{Eq:Liouville Equation}).
\begin{figure}
\includegraphics[width=\columnwidth]{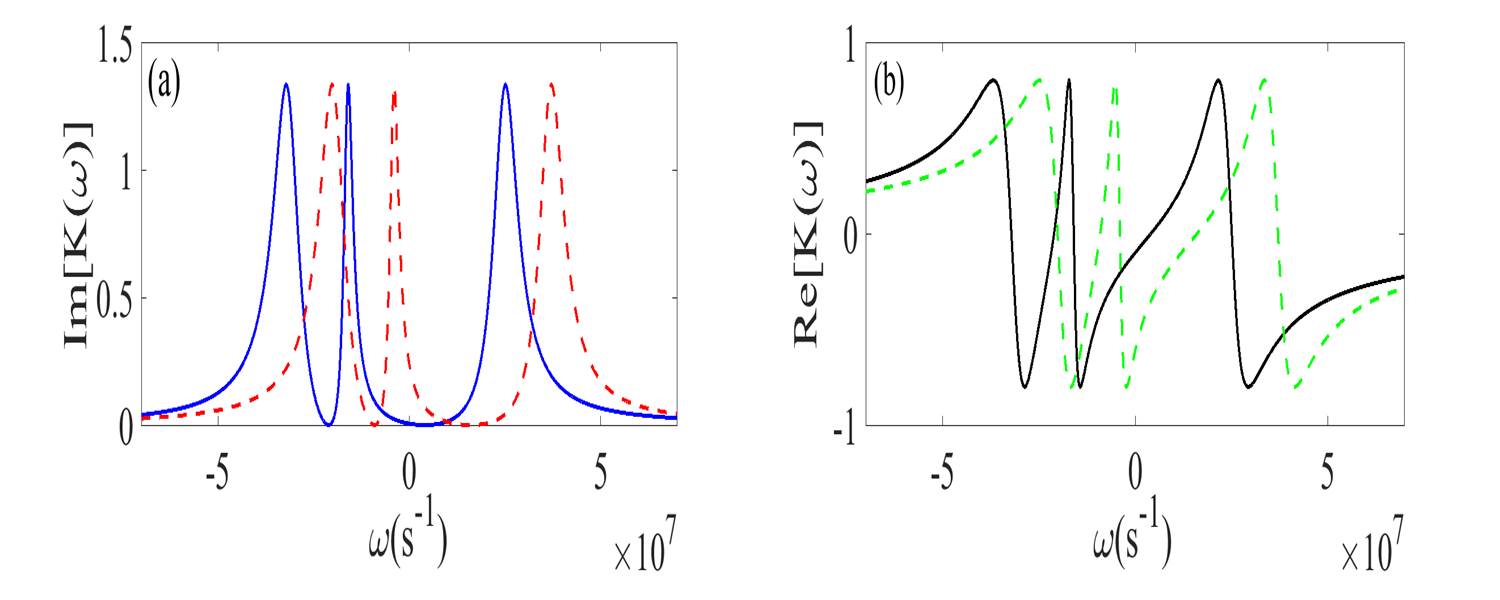}
\caption{\label{fig:onem}
Optical properties of the $\text{Pr}^{3+}$-ions in our hybrid waveguide.
Panel~(a) represents the absorption of the atomic medium for right~(blue solid-line) and left~(red dashed-line) circular polarizations. Panel~(b) depicts the chromatic dispersion of the SPWs for right~(black solid-line) and left~(green dashed-line) circularly polarizated components. Parameters are $|\Omega_\text{s}|=38~\text{MHz}$, $\Omega_\text{c}=80~\text{MHz}$, $\Delta_\text{s}=2~\text{MHz}$, $\Delta_{c}=0$, $\Delta_\mu=0.5~\text{MHz}$ and $\Delta_\text{p}^+=-\Delta_\text{p}^-=5~\text{MHz}$. The suitable temperature is $T=4~\text{K}$ and $\Omega_\mu\approx2.5~\text{MHz}$. Parameters related to NIMM layer are $\varepsilon_{\infty}=\mu_{\infty}=1.2$, $\omega_\text{e}=1.37\times10^{16}~\text{s}^{-1}$, $\omega_\text{m}=10^{15}~\text{s}^{-1}$, $\gamma_\text{e}=2.37\times10^{13}~\text{s}^{-1}$ and $\gamma_\text{m}=10^{12}~\text{s}^{-1}$.}
\end{figure}

\subsection{\label{Methods} Methods}
In this work, the coupled plasmonic dark rogue waves, phase deformation of plasmonic waves and controllable polaritonic frequency combs are obtained using pertubative method employing the Darboux transformation. In \S\ref{Multiple_scale} we discuss the multiple scale method for perturbative solution of the Maxwell-Liouville equations. In \S\ref{Lax-pair} we discuss the main steps in formation of Lax pairs and review the dressed Darboux transformation for solving the Manakov system of equations.

\subsubsection{\label{Multiple_scale} Multiple scaling variable and asymptotic expansion}
Our methods for solving Maxwell-Bloch equations in this polaritonic waveguide 
are based on the asymptotic expansion commensurate with the multiple-scale fast and slow variables.
Asymptotic expansion for an arbitrary function $f(x,t)$ with respect to sequences $\mathcal{F}(x,t)$ and a perturbation parameter $\varepsilon$ is a series or terms written as
\begin{equation}
    f(x,t)-\sum_{l=0}^{l=N}\delta(\varepsilon_{l})\mathcal{F}_{l}(x,t)=\mathcal{O}\{\mathcal{F}_n(x,t)\}.
\end{equation}
We define
\begin{equation}
    \delta(\varepsilon_{l}):=\varepsilon^{l},
\end{equation}
and asymptotically expand the density-matrix elements ($\tilde{\rho}_{ij}$) and probe field Rabi frequency ($\Omega_\text{p}$) to obtain perturbative solution of Maxwell-Bloch equation. In our analysis, we also assume the position to be slow two-scale variable~($x_{0,1,2}$) and time as a slower one-step scaled variable~($x_{0,1}$). Our truncated third-order perturbative solution (i.e. solution up to $l=3$) for both probe laser polarizations would yield a coupled nonlinear Schr\"odinger equations.

\subsubsection{\label{Lax-pair}Lax pair and dressed Darboux transformation}
We start this subsection by introducing the our method to solve the Manakov system of equations. 
We begin with a brief discussion of Lax-pair formation and then we explain the general properties of our dressed Darboux transformation method to solve Manakov like system.  
The general solution of the Manakov system can be obtained using the dressed Darboux transformation~\cite{PhysRevLett.113.034101}. We focus on generation and propagation of fundamental polaritonic dark rogue waves,
but this method is powerful beyond our needs here such as yielding solutions of
bright-dark and bright-bright rogue waves in a focusing Manakov system~\cite{chen2015vector}.

We employ this method to convert Eq.~(\ref{Main_Manakov})
to connect with the Lax pair of the linear eigenvalue problem~\cite{degasperis2009multicomponent}
\begin{align}
    \frac{\partial \psi}{\partial x}=X\psi,\;\;\; \frac{\partial \psi}{\partial \tau}
    =T\psi,
    \label{Main_ODEs}
\end{align}
with $\psi=\psi(x,t,k)$ a solution of the aforementioned ODEs.
Here~$k$ is a complex parameter, and $X$ and $T$ are $N\times N$ matrices depending on $(x,\tau,k)$.

We let $\psi^{(0)}$ be the the singular solution of Eq.~(\ref{Main_ODEs})
at zeroth order
\begin{align}
  \frac{\partial \psi^{(0)}}{\partial x}=X^{(0)}\psi^{(0)},\;\;\; \frac{\partial \psi^{(0)}}{\partial \tau}=T^{(0)}\psi^{(0)},  
\end{align}
then the higher order solution of Eqs.~(\ref{Main_ODEs}) is obtained by assuming $D(x,t,k):=\psi\psi^{(0)-1}$ or consequently
\begin{align}
    \psi=D(x,t,k)\psi^{(0)},
\end{align}
with
\begin{equation}
    D(x,t,k)=\bm{1}+\frac{\bm{\mathcal{M}}(x,t)}{k-k_\text{pole}},
\end{equation}
here $\bm{\mathcal{M}}$ is the residue matrix and $k_\text{pole}$ the pole position in the complex plain $k$. Evaluating $\bm{\mathcal{M}}$ matrix is challenging and can be calculated only for a few cases of nonlinear optical waves. We calculate this matrix for special cases of equal polaritonic field amplitude in Appendix~\ref{Derivation of dark coupler rogue waves}. In this work, we employ this transformation to obtain the surface polaritonic analogue of coupled dark rogue waves.

Now we show the emergence of SP phase singularities corresponding to phase deformation of a polaritonic plane wave due to field localization at a specific time and position~\cite{frisquet2016optical}.
This emergence leads to generation of multimode polaritonic frequency combs in our proposed scheme
shown in Fig.~\ref{fig:one}.
Excitation and propagation of the dark nonlinear-polaritonic waves, polaritonic phase singularities, and frequency combs are achieved by coupling the two orthogonal polarizations
of the probe laser to the $\ket1\leftrightarrow\ket2$ transition. The NIMM layer with low Ohmic loss enables low-loss SPW propagation.
Dual EIT windows emerging at the atom-NIMM interface enables us to modify nonlinearity and dispersion of the two-mode SPs.

\section{\label{results}Results}
In this section we discuss the main results of our study.
We explain the linear properties of excited two-mode SPWs, obtain the coupled NLSE system of equations and give a mathematical description to derive the Manakov system of equations in \S\ref{linearExcitation}.
In \S\ref{Coupled}, we discuss generation of coupled dark rogue waves, formation of surface polaritonic phase singularities and stable propagation of multimode polaritonic frequency combs. 

\subsection{\label{linearExcitation} Coupled surface polaritonic wave excitation and propagation in the linear regime}
Excitation and propagation of the coupled SP waves would depend on the dispersion and dissipation of the interface.
Generation efficiency and propagation length of the plasmonic waves are maximized if the Ohmic loss of the NIMM layer and the linear absorption of the atomic medium are minimized.
Consequently, dispersion and dissipation of the atomic medium in the linear regime are modified.
As a result, in order to stably propagating coupled SP waves, the linear optical properties of the hybrid plasmonic waveguide should be modified.
Therefore, in this section, we take advantage of the dispersion controllability of atomic medium around DEIT windows and employ the low-loss behavior of NIMM layer in the optical region to formulate stably propagated coupled SP waves in the linear regime.    

We provide a detailed quantitative description of our proposed waveguide by employing singular perturbations to Maxwell-Bloch equations~\cite{PhysRevA.73.020302}. We obtain the zeroth-order solution of this perturbative solution (i.e., the steady-state solution) achieved by setting $\Omega_\text{p}^{\pm}=0$ and $(\partial/\partial t)=0$ as
\begin{align}
   \tilde{\rho}_{11}^{+(0)}=\tilde{\rho}_{11}^{-(0)}=1,\;\; \tilde{\rho}_{ij}^{{(0)}\pm}=0 \; \text{for}\; ij\neq11. 
\end{align}
Our approach is then based on perturbative, asymptotic expansions with multiple scale position ($x$) and time ($t$) variables~\cite{PhysRevA.98.013825,asgarnezhad2017coupler}
\begin{align}
    x_l=\varepsilon^{l}x,\;t_l=\varepsilon^{l}t,
    \label{Eq:two}
\end{align}
for~$\varepsilon$ the perturbation-scale parameter
\begin{equation}
    \varepsilon:=\max\left\{\left|\frac{\Omega_\text{p}^\pm}{\Omega_\text{c}}\right|,\left|\frac{\Omega_\text{p}^\pm}{\Omega_\mu}\right|,\left|\frac{\Omega_\text{p}^\pm}{\Omega_\text{s}}\right|\right\}
\end{equation}
and~$l$ the perturbation order.

We obtain the linear excitation regime by setting $l=1$
and assuming that perturbation of the atomic states is weak so only the
$\ket{j}\leftrightarrow\ket1$; $j\in\{2,3,4\}$ coherence term contributes to evolution of the coupled SPWs.
In this case
\begin{equation}
    \Omega_\text{p}^{\pm}
    	=\varepsilon\Omega_\text{p}^{\pm1},
\end{equation}
and the density-matrix elements are
\begin{equation}
  \tilde{\rho}_{ij}^{\pm}=\tilde{\rho}_{ij}^{\pm(0)}+\varepsilon\tilde{\rho}_{ij}^{\pm1}.
  \label{Eq:Linear_Density_Matrix}
\end{equation}
We treat a probe field with pulse-envelope function ($F^{\pm}$) perturbed as
\begin{equation}
    \Omega_\text{p}^{\pm1}=F^{\pm}\exp\{\text{i}\vartheta^{\pm}\}
    \label{Eq:Linear_Rabi_Frequency}
\end{equation}
with $\vartheta^{\pm}=K^{\pm}(\omega)x-\omega t$; $\omega$ is the perturbation frequency of the SP waves and $K^{\pm}(\omega)$ is the chromatic linear dispersion of the circular polarization components. The first order density-matrix elements are perturbed as
\begin{equation}
    \tilde{\rho}_{j1}^{\pm1}=\zeta(z)F^{\pm}a_{j1}^{\pm1}\exp\{\text{i}\vartheta\},\;j\in\{2,3,4\}
\end{equation}
with other $\tilde{\rho}_{ji}^{\pm1}=0$. We evaluate the linear dispersion by mapping $\Delta_\text{p}\mapsto\Delta_\text{p}^{\pm}$ as
\begin{widetext}
\begin{equation}
    K^{\pm}(\omega)=\frac{\omega}{n_\text{eff}\text{c}}+\left\langle\kappa_{21}^{\pm}\frac{|\zeta(z)\Omega_\text{c}|^2+(d_{31}^{\pm}-\text{i}\omega)(d_{41}^{\pm}-\text{i}\omega))}{\Omega_\text{s}\Omega_\text{c}\bar{\Omega_\mu}^2-|\zeta(z)|^2[\Omega_\text{c}^2(\omega+\text{i}d_{21}^{\pm})+\Omega_\text{s}^2(\omega+\text{i}d_{41}^{\pm})]+(\omega+\text{i}d_{31}^{\pm})[(\omega+\text{i}d_{21}^\pm)(\omega+\text{i}d_{41}^\pm)-|\Omega_\text{m}|^2]}\right\rangle
\end{equation}
\end{widetext}
with
\begin{equation}
	\bar{\Omega}_\mu^2:=\zeta^2(z)\Omega_\mu+\text{c.c.}.
\end{equation}
The complete solution of the first-order perturbation is given in Appendix~\ref{Exact First order}.

Chromatic dispersion of the SPWs in the interface between the atomic medium and ultra-low loss metamaterial layer is shown in Fig.~\ref{fig:onem}. This figure depicts the formation of DEIT~\cite{PhysRevA.67.023811} and the modification of the frequency spectrum by employing small frequency shifts to the probe field polarization components. This spectral modification leads to absorption cancellation in the dispersion spectrum. Therefore, each polarization component ($\Omega_\text{p}^\pm$) excites stable polaritonic waves within generated transparency windows at different spectral frequencies that can be propagated through interface.

These generated spectral windows depend on the frequency separation of DEIT windows,
which can be effectively modified by finely tuning the microwave-field intensity with a radio-frequency generator. We achieve stable coupled-polaritonic waves propagation for
\begin{equation}
	\omega_+\approx9.3~\text{MHz};
	\omega_-\approx-8~\text{MHz}.
\end{equation}
In this case, the optical properties of the hybrid interface
\begin{equation}
    \varepsilon_{d}:=1+\chi(\omega)
\end{equation}
here $\chi(\omega)$ is the susceptibility of the interface is modified as
\begin{align}
    \text{Im}[\varepsilon_\text{d}]\ll1,\;\text{Re}[\varepsilon_\text{d}(\omega)]\leq1
\end{align}
leading to stable propagation of the SPWs. 

We employ the nonlinearity and higher-order dispersion in our analysis by taking advantage of the tightly confined coupled polaritonic waves. We incorporate the nonlinearity of the atomic medium by considering both the self-Kerr nonlinearity ($\chi_\text{pp}^{\imath\imath(3)}$) and the cross-Kerr nonlinearities ($\chi_\text{pp}^{\imath\jmath(3)}$); with $\imath,\jmath\in\{\pm\}$ and $\imath\neq\jmath$ as
\begin{align}
    \chi^+(\omega)=&\chi_\text{p}^{+(1)}(\omega)+\left|\Omega_\text{p}^+\right|^2\chi_\text{pp}^{++(3)}+\left|\Omega_\text{p}^-\right|^2\chi_\text{pp}^{+-(3)},\nonumber\\
    \chi^-(\omega)=&\chi_\text{p}^{-(1)}(\omega)+\left|\Omega_\text{p}^-\right|^2\chi_\text{pp}^{--(3)}+\left|\Omega_\text{p}^+\right|^2\chi_\text{pp}^{-+(3)},
\end{align}
with
\begin{equation}
    \chi_\text{p}^{\pm1}(\omega)=\frac{N_\text{a}|\bm{p}_{21}|a_{21}^{\pm1}}{\varepsilon_0|\bm{E}_\text{p}^\pm|},
\end{equation}
and we consider the higher order dispersion by assuming $\omega_{\pm}$ as a carrier frequency and employing Taylor expansion of $K(\omega)$ around the spectral transparency windows as
\begin{equation}
    K^\pm(\omega)=\sum_{m=0}^{\infty}\left(\frac{K_\text{m}^\pm}{m!}\right)(\omega-\omega_{\pm})^{m},
\end{equation}
with $K_\text{m}^\pm=[\partial^{m}K^\pm(\omega)/\partial\omega^{m}]_{\omega=\omega_{\pm}}$. 

\begin{figure}
\includegraphics[width=\columnwidth]{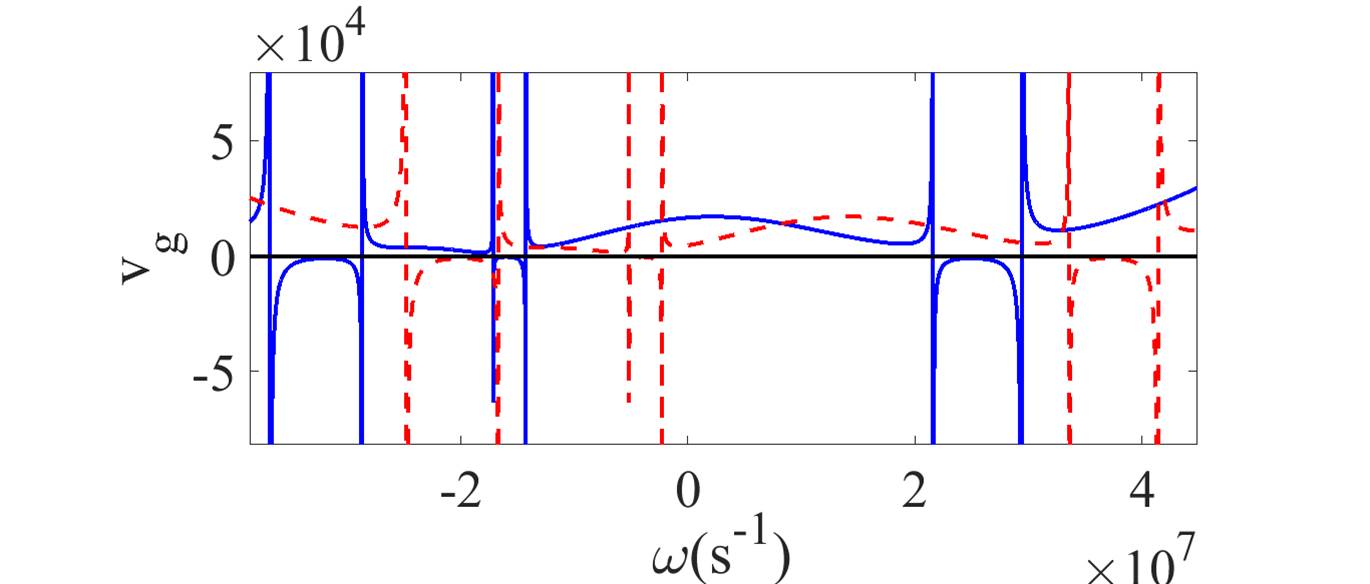}
\caption{\label{fig:twom}%
	Dynamics of group velocity for the two-mode surface-polaritonic waves. The blue solid line represents the linear group velocity of the SPWs excited by right circularly polarized while the red dashed line represents the SPP dynamics for the left circularly polarized light. The parameters used in this simulation are the same as the Fig.~\ref{fig:onem}. 
}
\end{figure}

Evolution of group velocity for two-mode SPWs in terms of perturbation frequency of the polaritonic waves is depicted in Fig.~\ref{fig:twom}.
Different group velocities for two-mode SPWs are evident,
and we see the slow-light propagation in spectral windows of DEIT. The small group velocity mismatch around the centre of spectral transparencies leads to the intensity profile overlap of the two-mode polaritonic modes which provides coupled-mode surface-polaritonic waves.
Therefore, points corresponding to $\omega_\pm$
are suitable for stable propagation of coupled surface-polaritonic waves.

\subsubsection{\label{nonlinearSPP} Excitation and propagation of coupled nonlinear SPWs}
Modifying the linear properties of the SPWs in the interface between the atomic medium and confining the electric field component of the polaritonic waves in our nonlinear hybrid waveguide yields the excitation and propagation of coupled two-mode nonlinear polaritonic waves. As our waveguide has the potential to control self-defocusing nonlinearity of the two-mode SPWs, we predict the excitation of surface polaritonic coupled dark wave formation, polaritonic phase singularities and controllable multimode frequency combs generation.

Therefore, we organize this section as follows: In \S\ref{derivation} we explore the realistic parameters to demonstrate that Dynamics of the two-mode SPs in our polaritonic waveguide is described by the Manakov system and in \S\ref{Coupled} we discuss the consequences of the two-mode nonlinear SPWs evolution in the hybrid polaritonic waveguide and predict the formation of surface polaritonic coupled dark rogue waves, polaritonic phase singularities and controllable surface polaritonic frequency combs.

\subsubsection{\label{derivation} Derivation of standard Manakov system}
This subsection describes the mathematical details of Manakov system derivation. First, we employ the multiple scale variable method and asymptotic expansion to derive the coupled NLSE.
Next we test the efficiency of our waveguide for a set of experimentally accessible parameters.
Finally, we give a detailed technical discussion to derive the coupled NLSE to the Manakov system of equations.

To evaluate the propagation of the SPWs in the nonlinear regime, we employ asymptotic expansions to probe field Rabi-frequency and density-matrix elements for both circular polarizations as
\begin{align}
    \Omega_\text{p}^\pm(\bm{r},t)=&\sum_l\varepsilon^{l}\Omega_\text{p}^{\pm(l)},\nonumber\\
    \tilde{\rho}_{ij}^\pm-\tilde{\rho}_{ij}^{\pm(0)}=&\sum_l\varepsilon^{l}\tilde{\rho}_{ij}^{\pm(l)},
\end{align}
with $\Omega_\text{p}^{\pm(l)}$ and $\tilde{\rho}_{ij}^{\pm(l)}$ the $l^\text{th}$-order perturbation of the density-matrix elements and probe field Rabi frequencies, respectively.
We obtain the second-order perturbative solution of the Maxwell-Liouville equations
by assuming
\begin{align}
    \tilde{\rho}_{jj}^{\pm2}=&a_{jj}^{\pm2}|\zeta(z)|^2|F^\pm|^2\exp\{\alpha x_2\},\nonumber\\
    \tilde{\rho}_{11}^{\pm2}=&-\left(\tilde{\rho}_{22}^{\pm2}+\tilde{\rho}_{33}^{\pm2}+\tilde{\rho}_{44}^{\pm2}\right),\nonumber\\
\end{align}
for diagonal matrix elements and
\begin{align}
    \tilde{\rho}_{23}=&a_{23}^{\pm2}|\zeta(z)|^2|F^\pm|^2\exp\{\alpha x_2\},\nonumber\\
    \tilde{\rho}_{24}=&a_{24}^{\pm2}|\zeta(z)|^2|F^\pm|^2\exp\{\alpha x_2\},\nonumber\\
    \tilde{\rho}_{34}=&a_{23}^{\pm2}|\zeta(z)|^2|F^\pm|^2\exp\{\alpha x_2\},\nonumber\\
    \tilde{\rho}_{j1}=&a_{23}^{\pm2}|\zeta(z)|^2\frac{\partial F^\pm}{\partial t_1}\exp\{\text{i}\vartheta^\pm\},
\end{align}
for nondiagonal density-matrix elements, with $j\in\{2,3,4\}$, and
\begin{equation}
	\alpha=\varepsilon^2\text{Im}[K(\omega)]
\end{equation}
representing the loss coefficient.

Now we obtain dynamics of the coupled nonlinear SPWs within hybrid interface. To this aim, we assume that the probe laser is a weak field and neglect higher-order perturbation ($\Omega_\text{p}^{\pm(l>1)}=0$).
Dynamics of the coupled SPWs in the second order approximation ($l=2$) is then expressed as
\begin{equation}
    \text{i}\left(\frac{\partial}{\partial x_1}+\frac{1}{v_\text{g}^\pm}\frac{\partial}{\partial t_1}\right)F^\pm=0,\,
    v_\text{g}^\pm=\left[\frac{\partial K(\omega)}{\partial\omega}\right]^{-1}_{\omega=\omega_{\pm}},
\end{equation}
implying that two-mode SPWs propagate with group velocities $v_\text{g}^\pm$ and with probe pulse-envelope function $F^\pm$. The pulse envelope depends on the nonlinearity and group-velocity dispersion of the interface.
The solvability condition ($\Omega_\text{p}^{\pm(3)}=0$) for the third order $l=3$ requires
\begin{equation}
   \text{i}\frac{\partial F^+}{\partial x_2}-\frac{K_2^+}{2}\frac{\partial F^+}{\partial t_1^2}-\left(W_{++}|F^+|^2+W_{-+}|F^-|^2\right)F^+=0
\end{equation}
and
\begin{equation}
   \text{i}\frac{\partial F^-}{\partial x_2}-\frac{K_2^-}{2}\frac{\partial F^-}{\partial t_1^2}-\left(W_{-+}|F^+|^2+W_{++}|F^-|^2\right)F^-=0,
   \label{Eq:Coupled NLSE equation.}
\end{equation}
with
\begin{equation}
    K_2^\pm=[\partial^2 K(\omega)/\partial\omega^2]|_{\omega=\omega_{\pm}}
    \label{Eq:Two_mode_GVD}
\end{equation}
characterizing atomic GVD.
Also
\begin{align}
    W_{\imath\imath}=&\frac{\hbar^2\omega_\text{p}^\imath }{2\text{c}|\bm{p}_{21}|^2}\chi_\text{pp}^{(3)}(-\omega_{\imath\imath},\omega_{\imath\imath},-\omega_{\imath\imath},\omega_{\imath\imath}),\nonumber\\
    W_{\imath\jmath}=&\frac{\hbar^2 \omega_\text{p}^\imath }{2\text{c}|\bm p_{21}|^2}\chi_\text{pp}^{(3)}(-\omega_{\imath\jmath},\omega_{\imath\jmath},-\omega_{\imath\jmath},\omega_{\imath\jmath}),
\end{align}
denotes self-phase modulation~(SPM) $W_{++(--)}$ and cross-phase modulation~(XPM) $W_{+-(-+)}$ of the atomic medium, which are calculated using
\begin{widetext}
\begin{equation}
    \chi_\text{pp}^{\imath\imath(3)}=\left\langle\zeta(z)|\zeta(z)|^2\frac{D_{11}(a_{11}^{\imath\imath(2)}-a_{22}^{\imath\imath(2)})+(\zeta^{*2}(z)\Omega_\text{s}\Omega_\text{c}+\Omega_\mu^*(\text{i}d_{31}+\omega))a_{32}^{\imath\imath(2)})+D_{42}a_{42}^{\imath\imath(2)})}{\Omega_\text{s}\Omega_\text{c}\bar{\Omega_\mu}^2-|\zeta(z)|^2[\Omega_\text{c}^2(\omega+\text{i}d_{21}^\pm)+\Omega_\text{s}^2(\omega+\text{i}d_{41}^\pm)]+(\omega+\text{i}d_{31}^\pm)[(\omega+\text{i}d_{21}^\pm)(\omega+\text{i}d_{41}^\pm)-|\Omega_\mu|^2]}\right\rangle
\end{equation}
\end{widetext}
with
\begin{equation}
   D_{11}:=|\zeta(z)\Omega_\text{c}|^2+(d_{31}^\pm-\text{i}\omega)(d_{41}^\pm-\text{i}\omega)
\end{equation}
and
\begin{equation}
   D_{42}:=\zeta(z)\Omega_\text{c}\Omega_\mu^*+\text{i}\zeta^*(z)\Omega_\text{s}(d_{41}+\text{i}\omega).
\end{equation}
Eqs.~(\ref{Eq:Coupled NLSE equation.}) and~(\ref{Eq:Two_mode_GVD}) have GVD, SPM and XPM as complex coefficients and are referred to as coupled Ginsburg-Landau equations.

In obtaining Eqs.~(\ref{Eq:Coupled NLSE equation.}) and~(\ref{Eq:Two_mode_GVD}) we neglect group-velocity dispersion~(GVD) of the NIMM layer,
which is $10^{-5}\times$ GVD of the N-type atomic medium. Furthermore, only $x_{0,1,2}$ and $t_{0,1}$ matter because one can neglect
(i)~the second derivative of $x_1$ due to the slowly varying amplitude approximation and (ii)~higher-order time scales $t_{l>1}$ and $x_{l>2}$ due to negligible higher-order dispersion effects. Our approach exploits controllable (DEIT) windows with separation frequency~$\delta\omega_*$
and low Ohmic loss of the NIMM layer to excite two-mode linear and nonlinear SPW modes.

We treat SPs as two-mode plane waves with total energy~$\mathcal{E}$. 
Furthermore,
we assume that excited nonlinear SPs have wide initial temporal pulse width 
\begin{equation}
  \tau_\text{p}^+\approx\tau_\text{p}^-
    :=\tau_\text{p}  
\end{equation}
and half Rabi frequency
\begin{equation}
    U_0
        :=\sqrt{\frac{|K_2^-|}{2\tau_\text{p}^2
            |W_{--}|}}.
\end{equation}

Plane SPWs generated by the probe laser with peak power $P_0$ can propagate up to several nonlinear length units
\begin{equation}
   L_\text{N}
    :=\frac{1}{U_0|W_{--}|}
\end{equation}
in the low-atomic absorption limit $\alpha\ll1$ if imaginary parts of SPM and GVD are much lower than real parts.

Effective group velocity, GVD and the resultant SP drift are~\cite{PhysRevLett.82.2661}
\begin{equation}
    v^\pm
        =\frac{2v_\text{g}^+v_\text{g}^-}{v_\text{g}^-\pm v_\text{g}^+},\,
    \Delta v_\text{g}=v^+-v^-,\,
    \mathcal{D}=-\frac{\tau_\text{p}\Delta v_\text{g}} {K_2^-v_\text{g}^2},
    \label{eq:drift}
\end{equation}
respectively and
average group-velocity mismatch is
\begin{equation}
    \bar{v}:=\pi\delta\omega_*\tau_0.
\end{equation}
Dispersion length and group-velocity mismatch length are
\begin{align}
    L_\text{D}:=\frac{2\tau_\text{p}^2}{|K_2^-|},\;L_\delta:=\frac{\tau_\text{p}}{|v^-|},
\end{align}
respectively, whence normalized effective group velocity is
\begin{equation}
    g_\text{d}:=\text{sgn}(\delta)\frac{L_\text{D}}{L_\delta}.
\end{equation}
Normalizing GVD $K_2^\pm$, SPM $W_{++(--)}$ and XPM $W_{\pm(\mp)}$ according to
\begin{align}
    g_\text{D}^\pm:=\frac{K_2^\pm}{|K_2^-|},\; g_\text{N}^{(\imath\jmath)}:=\frac{W_{\imath\jmath}}{W_{--}},
\end{align}
for $g_\text{D}^\pm$ and $g_\text{N}^{(\imath\jmath)}$, with $\imath,\jmath\in\{\pm\}$, denoting normalized GVD, SPM and XPM, respectively.

We assume
\begin{align}
   s=x/L_\text{D},\,\tau:=t-x/v^+,\,
   \sigma
    :=\sqrt{2}\tau/\tau_0
\end{align}
and employ the mapping
\begin{equation}
    \text{i}\left(\frac{\partial}{\partial s}\pm g_\text{d}\frac{\partial}{\partial\sigma}\right)
    \mapsto\frac{\partial}{\partial\mathcal{S}^\pm},
\end{equation}
and use
\begin{equation}
    u^\pm:=\left(\frac{\Omega_\text{p}^\pm}{U_0}\right)\exp\{-\bar{\alpha}x\}
\end{equation}
to see that the two-mode normalized SP wave dynamics is described by the coupled nonlinear Schr\"odinger equation~\cite{PhysRevLett.113.034101}
\begin{equation}
    \frac{\partial u^\imath }{\partial \mathcal{S}^\imath }-\frac{g_{\text{D}}^\imath }{2}
    \frac{\partial^2u^\imath}{\partial\sigma^2}
    -\left(g_\text{N}^{\imath+}|u^+|^2
    +g_\text{N}^{\imath-}|u^-|^2\right)u^\imath
    =0,
    \label{Eq:manakov}
\end{equation}
which can be transformed to the coupled nonlinear Schr\"odinger equation under certain parameters of the hybrid waveguide.

Nonlinear polarization ($\bm{P}_\text{NL}$) of the interface 
evidently effects SPW propagation via small perturbations on linear SPs. The perturbed wave has amplitude $p,q\ll1$ and frequency modulation~$\Omega_\pm$ with modulation parameter $\kappa_\pm$ and initial instability frequency ($\nu_\text{mod}$).
Due to nonlinearity of the medium, the wave experiences chirping
\begin{equation}
    \mathcal{C}^\pm=\frac{\partial\left[\arg\left(\Omega_\text{p}^{\text{D}\pm}\right)\right]}{\partial t},
\end{equation}
leading to coupled-frequency-comb generation.
Nonlinear SPs arise by assuming~$\xi^\pm$ 
has a slowly varying amplitude modification to the linear SPs.
\begin{figure}
\includegraphics[width=\columnwidth]{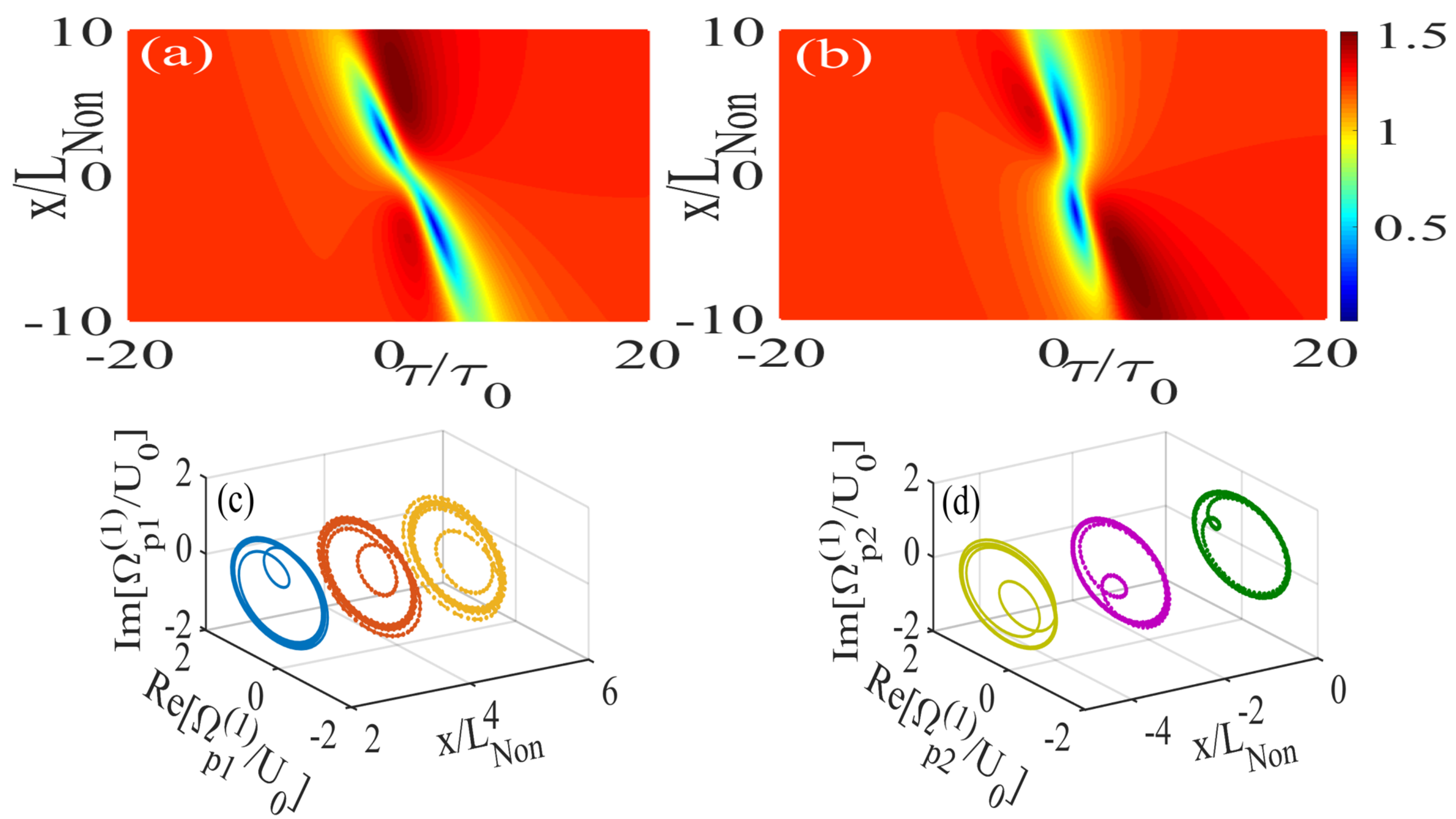}
\caption{%
Formation of first-order dark SProgue wave by coupling two plane wave SPs.
Panels~(a,b) show intensity patterns for plane-wave SPs.
The blue-color map shows zero-intensity points for the coupled SPs whereas the dark-red color map depicts the enhanced intensity of SPWs. Panels~(c,d) show the evolution of the coupled SPWs depicted in panel~(a) in the complex $\text{Im}[\Omega_\text{p}^{(\text{D}+)}]-\text{Re}[\Omega_\text{p}^{(\text{D}+)}]$ plane. Plots are obtained for  atomic absorptions $\alpha^+\approx\alpha^-=0.15$, $|\Omega_\mu|=60~\text{kHz}$, and $\delta\omega_*=20~\text{MHz}$. We use $\tau=-3.57\tau_\text{p}$ and $\tau=0.28\tau_\text{p}$ for panel~(c) and panel~(d), respectively.}\label{fig:two}
\end{figure}

We numerically analyse performance of our nonlinear polaritonic waveguide using realistic parameters for both the atomic medium~\cite{wang2008all}
and the NIMM layer~\cite{xiao2010loss,kamli2011quantum}. Radiative decay is $\Gamma_{21}^{\text{R}}
    =\Gamma_{23}^{\text{R}}=9~\text{kHz}$,
nonradiative decay is  $\Gamma_{42}^{\text{NR}}
    =9~\text{kHz}$,
$\Gamma_{31}^{\text{NR}}=10~\text{kHz}$,
and atomic density is $N_\text{a}=4.7\times10^{18}~\text{cm}^{-3}$. We assume $\left|\Omega_\mu\right|\approx2~\text{MHz}$,
$\Delta_\mu=0.5~\text{MHz}$ and $\theta=\pi/2$.
Signal and coupling light fields have frequencies $\Omega_\text{s}=37.70~\text{MHz}$, $\Omega_\text{c}\approx80~\text{MHz}$ and detunings $\Delta_{s}\approx2~\text{MHz}$, $\Delta_\text{c}=0$ from atomic transitions. These light fields can be prepared from a laser using acousto-optic modulation. Under these conditions, we achieve controllable atomic DEIT windows at
\begin{align}
    \omega_-\approx-16.19~\text{MHz},\;\omega_+\approx10.30~\text{MHz}.
\end{align}
DEIT-window frequency separation is $\delta\omega_*:=\omega_+-\omega_-$. Coherent excitation and stable coupled-SPW propagation  is achieved by adjusting probe frequency within DEIT windows.

We find:
(i)~For the DEIT window centreed at $\omega^-$, nonlinear SPs propagate with 
\begin{equation}
    v_\text{g}^-=1.23\times 10^{-5}\text{c},
\end{equation}
$W_{--}\approx W_{-+}$, and
\begin{align}
    K_2^-=&(3.29+0.20\text{i})\times10^{-15}~\text{cm}^{-1}\text{s}^2,\nonumber\\
    W_{--}=&(-1.59+0.05\text{i})\times10^{-14}~\text{cm}^{-1}\text{s}^2.
\end{align}
(ii)~For the DEIT window centreed at $\omega^+$,
nonlinear SPs propagate with
\begin{equation}
    v_\text{g}^+=1.896\times 10^{-5}\text{c},
\end{equation}
we have $W_{+-}\approx W_{++}$, and
\begin{align}
  K_2^+=&(2.90+0.10\text{i})\times10^{-15}~\text{cm}^{-1}\text{s}^2,\nonumber\\ 
  W_{++}=&(-1.46+0.30\text{i})\times10^{-14}~\text{cm}^{-1}\text{s}^2.
\end{align}
Cases (i) and (ii) imply
\begin{align}
   g_\text{N}^{(\imath\jmath)}\approx1,\; g_\text{D}^+\approx g_\text{D}^-=1,\; g_\text{d}\mapsto0,
\end{align}
and $v^-\approx0$,
so Eq.~(\ref{Eq:manakov}) transforms to standard Manakov equations~\cite{frisquet2016optical}, which possess dark coupled-polaritonic nonlinear waves~(see Appendix~\ref{Derivation of dark coupler rogue waves} for more details).

\subsection{\label{Coupled} Coupled surface polaritonic dark rogue wave excitation, polaritonic phase singularities and controllable polaritonic frequency combs formation}
 Our proposed hybrid waveguide supports coupled dark SP rogue waves excited by amplitude and detuning modulation of driving fields shown in Figs.~\ref{fig:two}(a,b),
 which depicts SP dynamics.
 Coupled dark rogue waves emerge due to modulation instability and resultant nonlinear interference of perturbed two-mode plane SPWs.
 By controlling nonlinearity and dispersion commensurate with polarization-modulation instability,
 we obtain SP-propagation dynamics at the atom-NIMM interface shown in Figs.~\ref{fig:two}(c,d) 
 depicting polaritonic trajectories
 (dynamics in the $\text{Im}[\Omega_{\text{p}\imath}^\imath]-\text{Re}[\Omega_{\text{p}\imath}^\imath]$ plane).
Dark rogue waves emerge at points corresponding to \begin{equation}
    \left(\text{Re}[\Omega_{\text{p}\imath}^\imath],\text{Im}[\Omega_{\text{p}\imath}^\imath]\right)
    =(0,0)
\end{equation}
(`zero-intensity points') of the SP trajectory map.

Formation of coupled polaritonic-dark rogue waves 
could enable multimode high-speed polaritonic switches~\cite{PhysRevA.99.051802} and a phase modulators~\cite{melikyan2014high}.
To explore these possibilities, we let
\begin{equation}
   \Omega_\text{p}^{\text{D}\pm}
   =\left|\Omega_\text{p}^{\text{D}\pm}|\right|
   \exp\left[\text{i}\arg\left(\Omega_\text{p}^{\text{D}\pm}\right)
   \right], 
\end{equation}
and consider small frequency separation
\begin{align}
\frac{\delta\omega_*}{\omega_\text{p}}\ll1,\; |\Omega_\text{p}^+|\approx|\Omega_\text{p}^-|.  
\end{align}
Using reasonable parameters,
we obtain
\begin{equation}
   \arg\left(\Omega_\text{p}^{\text{D}+}\right)
   \approx\arg(\Omega_\text{p}^{\text{D}-})\approx\pi 
\end{equation}
which is the phase shift between the initial and recovered state of the two-mode SPWs. This phase shift is also useful for multimode SP phase modulation. As our polaritonic waveguide exhibits highly defocusing nonlinear dispersion, significantly compressed dark temporal pulse through grow-return cycle of perturbed SPP wave is expected, which yields a fast~(switching time falling to $10^{-8}\tau_0$) multimode-SP switching. 
\begin{figure}
\includegraphics[width=\columnwidth]{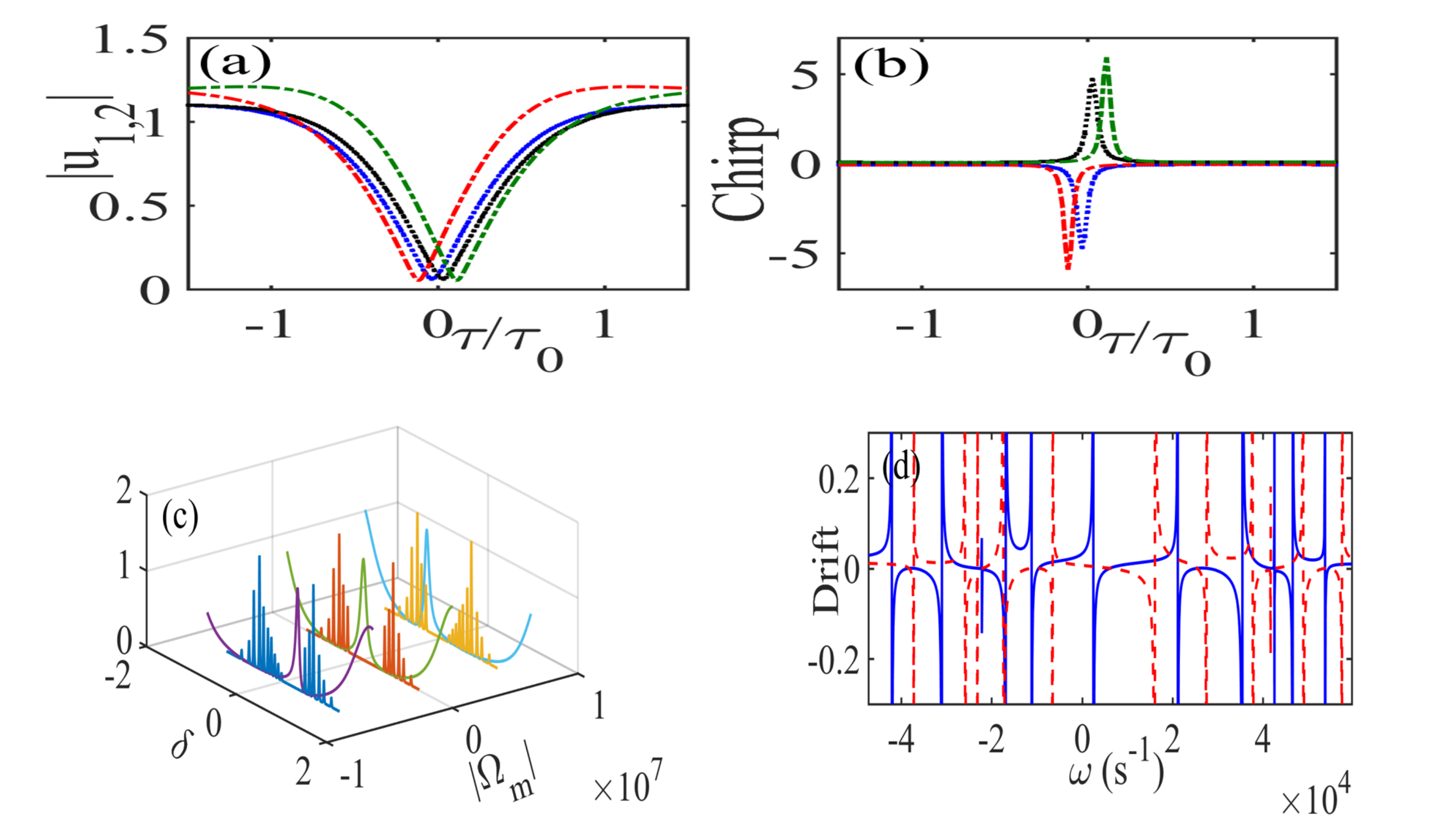}
\caption{Impact of symmetry properties
of the dual EIT windows on SP dark rogue waves: (a) The evolution of nonlinear SPs for $\phi^+=0$, $\phi^-=\pi/2$, $\varepsilon^+=\varepsilon^-\approx 0.04$ and $\nu_\text{mod}=\delta\omega_*/3$. Black and blue solid lines are obtained for $\delta\omega_*=10~\text{MHz}$ and green and red dotted lines are obtained for $\delta\omega_*=30~\text{MHz}$ (b) Time evolution of the two-mode SP phase as a function of the frequency splitting of the EIT windows.
(c) SP frequency combs with symmetric or asymmetric shapes by varying the Rabi frequency of the $\mu$-wave driving field. (d) SP Drifts~(\ref{eq:drift}).}
\label{fig:three}
\end{figure}

Zero-intensity points of two-mode SPs
(spatial positions of polaritonic dark rogue-wave) depend on EIT-window frequency separation,
which we illustrate by letting
$\mu$w field $|\Omega_\mu|\ll \Omega_\text{p}$,
initial power $P_0=10\mu\text{W}$ for SP generation, and dual EIT windows with $\theta_\text{m}=\pi/2$, and using $\delta\omega_*$ as the control parameter.
The dual EIT windows become symmetric for resonance conditions $\omega\approx0$ and $K_j^+\approx K_j^-$,  
$j\in\{0,1,2\}$.

Using realistic parameter values, 
the intensity hole bifurcates leading to SP phase-singularity formation at
\begin{align}
 (-3.57\tau_\text{p},3.57L_\text{N}),\;  (0.28\tau_\text{p},-2.60L_\text{N})   
\end{align}
for left-circular and
\begin{align}
 (-1.65\tau_\text{p},2.62L_\text{N}),\; (-0.93\tau_\text{p},-3.57L_\text{N})   
\end{align}
for right-circular polarization.

We numerically solve the Manakov-like system by assuming a small perturbation
\begin{align}
    u_0^\pm(t)
        :=&\sqrt{P_0}[1+\varepsilon^\pm\cos(2\pi\nu_\text{mod}t)]\text{e}^{\text{i}\phi^\pm}
    \label{Eq:initial}
\end{align}
in terms of modulation amplitude and modulation frequency.
We demonstrate bifurcation of coupled polaritonic dark rogue waves in time domain by modifying frequency splitting of the centre of symmetric dual EIT windows as seen clearly in Fig.~\ref{fig:three}(a).
The zero-intensity point bifurcates as we increase frequency separation of the EIT windows from $\delta\omega_*\approx10\text{MHz}$ to $\delta\omega_*\approx30\text{MHz}$. 

DEIT-window frequency separation produces symmetric and tunable frequency chirping as shown in Fig.~\ref{fig:three}(b),
achieved by numerically evolving the two-mode polaritonic phase.
Generating a symmetric chirp leads to multimode surface polaritonic frequency combs,
illustrated in Fig.~\ref{fig:three}(c) by numerically solving the Manakov-system for two input plane-wave SP fields with initial condition~(\ref{Eq:initial}).

Formation of symmetric DEIT windows leads to symmetric frequency chirps,
which are time-reversed quasi-zero SP-wave intensities 
\begin{align}
   \bm{\Theta}|u_1|=|u_2|,\;  \bm{\Theta}^{-1}|u_2|=|u_1| 
\end{align}
with~$\Theta$ denoting the time-reversal operator leading to formation of imitative two-mode polaritonic-frequency combs. 

With the $\mu$-wave field Rabi frequency as another control parameter (by increasing $|\Omega_\mu|$) in the case of asymmetric dual EIT windows, we control DEIT-window width as shown in Fig.~\ref{fig:three}(c). In this case, we expand the frequency-comb mode number ($\nu^\pm$) around the EIT-window central frequencies ($\omega_\pm$) as a power series of asymmetric SPW dispersion ($\{\mathcal{K}_l^\pm\}$~\cite{kippenberg2011microresonator} and neglect higher-order dispersion ($\{\mathcal{K}_{l>2}^\pm\}$),
yielding
\begin{equation}
  \omega^\pm=\omega_*^\pm+\mathcal{K}_1^\pm(x)\nu^\pm+\frac{\mathcal{K}_2^\pm(x)}{2}\nu^{\pm2}+\cdots.
  \label{eq:conditionfrequencycomb}
\end{equation}
Some polaritonic-frequency comb modes are absorbed
into the narrow EIT window due to different linear dispersions
($\mathcal{K}_1^+\neq\mathcal{K}_1^-$ and $K^+(\delta^+)\neq K^-(\delta^-)$).
This absorption leads to different frequency chirps and an asymmetric pattern for polaritonic frequency combs. 

Polaritonic frequency combs with sideband spacing~$\Delta$
fall within EIT windows,
and $\mathcal{N}^\pm:=\Delta/\delta^\pm$ polaritonic modes are excited at the atom-NIMM interface.
Increasing nonlinearity and suppressing higher-order dispersion
\begin{equation}
   |\mathcal{K}_2^\pm(x)|\ll\text{c}|\mathcal{K}_1^{\pm}(x)|^2 
\end{equation}
through the atom-NIMM interface, we achieve efficient polaritonic frequency combs in the broad EIT window, whereas,
in the narrow EIT window,
generated SP frequency combs are absorbed. 
\begin{figure}
\includegraphics[width=\columnwidth]{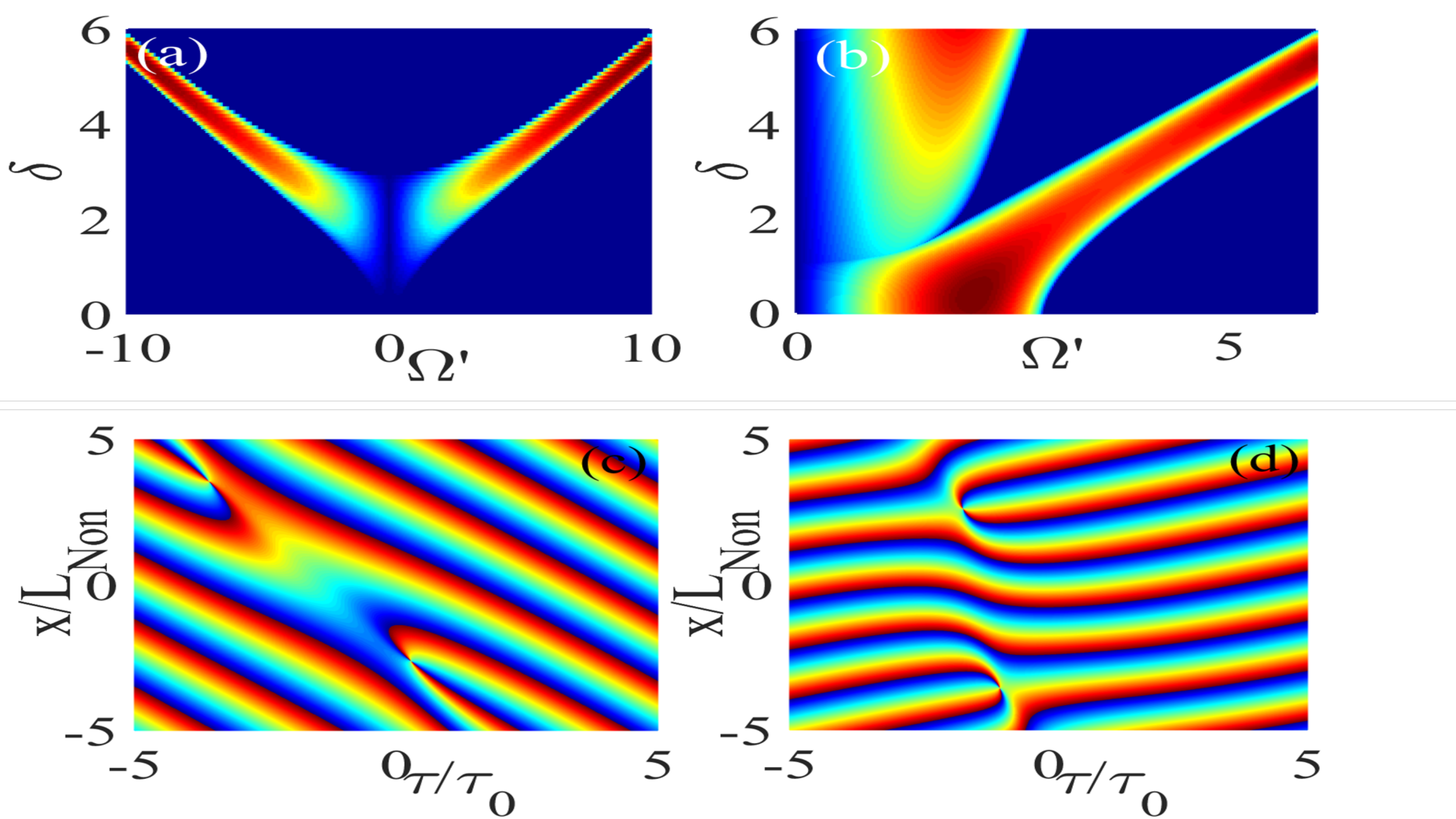}
\caption{%
Intensity map of the weakly perturbed coupled SP waves in the
(a)~normal and (b)~anomalous dispersion  regimes in the presence of giant self-defocusing nonlinearity.
Panels~(c,d) show phase evolution of the perturbed~$u^\pm$ SPWs at the atom-NIMM interface, respectively.
Phase singularity and consequent twisting of SPs at the position of stable polaritonic dark rogue waves are clearly seen.}
\label{fig:four}
\end{figure}

Now we explain the physical origin of polaritonic dark rogue-waves in our scheme using the SP-wave gain map and by considering the seeded polaritonic modulation instability. Thus, we assume small perturbations
\begin{equation}
u^\pm(x,t)=u_0^\pm\left[1+p^\pm\text{e}^{-\text{i}\Omega^\pm(\kappa^\pm \tilde{x}-\tau)}
        +q^{* \pm}\text{e}^{\text{i}\Omega^\pm(\kappa^{* \pm} \tilde{x}-\tau)}\right]
\end{equation}
by expanding SP dispersion and normalize nonlinear coefficients, detuning frequencies
\begin{equation}
 \delta^-\approx \delta^+:=\delta   
\end{equation}
and frequency modulation~$\Omega'$. We assume
\begin{equation}
  |\omega_*^-|\approx|\omega_*^+|,\;\Gamma^\pm=v_\text{g}^\pm/\Delta v_\text{g}  
\end{equation}
and map
\begin{equation}
  \mu\mapsto(2\kappa/g_\text{D})-2\delta,\; k+2\tilde{K}_0/g_\text{D}\mapsto\zeta.
\end{equation}
Then we define
\begin{equation}
  \tilde{K}_0:=K_0+1/(2L_\text{N}). 
\end{equation}
Linearizing for weak perturbation~\cite{chen2017versatile} and assuming 
\begin{equation}
    u_0^+\approx u_0^-
\end{equation}
yields
\begin{equation}
  \frac{1}{4}\left[\mu^2+2\frac{K_1\delta}{g_\text{D}}+\left(1+2\frac{K_2}{g_\text{D}}\right)\delta^2+\zeta\right]^2-\mu^2\delta^2-\text{A}^2=0  
\end{equation}
for
\begin{equation}
    \text{A}:=\left[k^-+K^-(\delta)\right]/g_\text{D}+\delta^2/2.
\end{equation}
The gain map for the anomalous-dispersion case ($g_\text{D}\approx1$) leads to excitation of nonlinear SP modes as polaritonic dark rogue waves commensurate with symmetric normalized DEIT windows
\begin{align}
   0.1<\delta<1.6,\; 0.5<\Omega'<2.4,
\end{align}
which create an allowed dispersion band (termed `base-band') for polaritonic modulation instability shown in Fig.~\ref{fig:four}(a).

For normal dispersion (self-focusing nonlinearity), maximum gain is achieved in the larger transparency window commensurate with the coupling laser intensity through Autler-Townes splitting. For a coupling laser with $\Omega_\text{c}\approx100~\text{MHz}$, $\theta_\text{m}=\pi/2$, observed base-band modulation instability is within
\begin{align}
  1.5<\delta<3,\; 0.3<\Omega'<1.4  
\end{align}
as shown in Fig.~\ref{fig:four}(b).

Our interacting multimode nonlinear SPWs arise by perturbing coupled Maxwell equations
so the total polarization of our hybrid plasmonic waveguide is linear and thus obtain linear electric fields
$\bm{E}_\text{SP}^\pm$
for each SP wave~(see Appendix~\ref{DerivationCoupled} for a detailed derivation of coupled Maxwell equation with nonlinear polarization).
By introducing nonlinear polarization and accounting for SPM and XPM as small perturbations,
the nonlinear electric field is modified to
\begin{equation}
    \nabla \xi^\pm
    =-\text{i}\frac{\omega^2\mu_0}{2k_\text{p}n_\text{eff}}
    \frac{\int\, \text{d}y\bm{P}_\text{NL}^\pm\bm{E}_\text{SP}^\pm}
    {\iint\,\text{d}y |\bm{E}|^2}.
\end{equation}
We achieve nonlinear SPWs by substituting this modification into coupled Maxwell equations and solve the resultant equation using 
\begin{equation}
   \xi^\pm=|\xi^\pm|\exp\left[\text{i}\arg(\xi^\pm)\right], 
\end{equation}
and
\begin{equation}
   g_\delta\approx g_\text{D}\mapsto0 
\end{equation}
in the weak perturbation limit.
Then
\begin{equation}
  \bm{E}=\xi^+\bm{E}_\text{SP}^++\xi^-\bm{E}_\text{SP}^-  
\end{equation}
in the total nonlinear electric field leading to an interference pattern. Consequently, we achieve a phase singularity in the spatial position due to destructive interference.

\section{\label{Discussion}Discussion}
We have presented a waveguide configuration to overcome the limitation in the controllability of coupled surface polariton's nonlinear interaction. We have employed both polarization and baseband modulation instability formalism in our waveguide configuration to generate the stable propagation of coupled dark rogue wave. Our theory incorporates the atomic dissipation and dispersion to the coupled SPWs for generating and controlling surface polaritonic phase singularities and multi-mode surface polaritonic frequency combs. 

Polaritonic dark rogue-wave formation depends on the group-velocity mismatch for the two polaritonic modes, as shown in Fig.~\ref{fig:three}(d) by simulating drift of the two SP modes. If XPM and SPM differ,
\begin{equation}
   g_\text{N}^{(\pm)}\approx g_\text{N}^{(\imath\imath)}+\delta g, 
\end{equation}
and the standard Manakov system becomes depleted, our predicted nonlinear waves cannot be observed under those conditions. In our scheme, stable polaritonic dark rogue waves are excited around the DEIT-window centre for which
\begin{equation}
\left|\mathcal{K}_2^{\pm}(x)\right|\ll\text{c}|\mathcal{K}_1^{\pm}(x)|^2\ll1    
\end{equation}
and the two-mode SPs have the same drift.

Coupled polaritonic dark rogue waves can also be generated and propagated in our hybrid plasmonic waveguide by adding the positive coherent nonlinear term~\cite{zhang2018rogue}
corresponding to energy exchange between coupled nonlinear SPWs. The energy-exchange effect between coupled SPWs in dark rogue wave formation in a plasmonic system is challenging and its inclusion goes beyond the scope of this work.
Moreover, higher-order surface polaritonic dark rogue waves can be excited and propagate in our nonlinear waveguide, however they can be assumed as a nonlinear superposition of a fixed well-prescribed number of the fundamental dark rogue waves~\cite{chen2015vector}. Our work only deals with the first-order polaritonic dark rogue waves.

Our nonlinear waveguide could serve as an SP phase rotor due to the emergence of coupled polaritonic dark rogue waves and resultant interference patterns shown in Fig.~\ref{fig:four}(c,d). To this aim, we add a small perturbation to the two-mode plane SPWs~(\ref{Eq:initial}) and consider their nonlinear interference during their propagation along the atom-NIMM interface. The SP twisted phase at the position of the polaritonic coupled dark rogue waves leads to singularity formation shown in Figs.~\ref{fig:four}(c,d), which are obtained by simulating the nonlinear dynamics of plane wave SPs at the nonlinear interface.

\section{\label{conclusions}Conclusions}
In summary, we propose a multimode polaritonic waveguide that exploits self-defocusing nonlinearity from N-type atoms
above the negative-index metamaterial
to control and excite strongly localized dark polaritonic waves.
We have shown generation and stable propagation of polaritonic dark rogue waves in this system for appropriate driving field intensities and detunings.
Moreover, we establish symmetric and anti-symmetric multimode SP frequency combs by modulating dual EIT windows commensurate with base-band SP-modulation instabilities. Our proposed waveguide twists the SP phase by nonlinear phase interference through the formation of dark-polaritonic rogue waves within symmetric atomic DEIT windows.~Energy exchange between the coupled SPWs would be interesting;
however, modeling energy interchange between these nonlinear plasmonic waves is challenging.

Our work focuses on fundamental plasmonic dark rogue waves;
higher-order coupled polaritonic rogue waves can be assumed to be a nonlinear superposition of a fixed well prescribed number of the fundamental dark rogue wave.
Therefore, our configuration could serve as a fast surface polaritonic modulator, multimode SP phase modulator and SP phase rotator, which could open prospects for investigating phase singularities in quantum-communication applications and for building compact nano-plasmonic devices.
\acknowledgments
BCS acknowledges support from NSERC and from the Alberta Government.
BK acknowledges financial support of the French Investissements d’Avenir program (PIA2/ ISITE-BFC, Contract ANR-15-IDEX-03, Project ``Breathing Light'').
\bibliography{ref}
\appendix
\begin{widetext}
\section{\label{Liouville}Solution of the Liouville equation}
The explicit solution of the Liouville operator is
\begin{align}
    \text{i}\left(\frac{\partial}{\partial t}+\Gamma_{21}\right)\tilde{\rho}_{22}+\text{i}\Gamma_{32}\tilde{\rho}_{33}-\text{i}\Gamma_{42}\tilde{\rho}_{44}+\frac12\left[\zeta(z)\Omega_\text{p}\tilde{\rho}_{12}+\zeta^*(z)\Omega_\text{s}\tilde{\rho}_{23}+\Omega_\text{m}^*\tilde{\rho}_{24}-\text{c.c.}\right]=&0,
    \end{align}
    \begin{align}
    \text{i}\left(\frac{\partial}{\partial t}+\Gamma_{33}\right)\tilde{\rho}_{33}-\text{i}\Gamma_{43}\tilde{\rho}_{44}+\frac12\left[\zeta(z)\Omega_\text{s}\tilde{\rho}_{32}-\zeta^*(z)\Omega_\text{s}\tilde{\rho}_{23}+\zeta(z)\Omega_\text{c}\tilde{\rho}_{34}-\zeta^*(z)\Omega_\text{c}\tilde{\rho}_{43}\right]=&0,
    \end{align}
    \begin{align}
    \text{i}\left(\frac{\partial}{\partial t}+\Gamma_{44}\right)\tilde{\rho}_{44}+\frac12\left[\zeta(z)\tilde{\rho}_{14}-\zeta^*(z)\tilde{\rho}_{41}+\Omega_\text{m}\tilde{\rho}_{42}+\Omega_\text{m}^*\tilde{\rho}_{24}\right]=&0,
    \end{align}
    \begin{align}
    \text{i}\left(\frac{\partial}{\partial t}+d_{21}\right)\tilde{\rho}_{21}+\frac12\zeta(z)\Omega_\text{s}\tilde{\rho}_{31}+\frac12\Omega_\text{m}\tilde{\rho}_{41}+\frac12\zeta^*(z)\Omega_\text{p}(\tilde{\rho}_{11}-\tilde{\rho}_{22})=&0,
    \end{align}
    \begin{align}
    \text{i}\left(\frac{\partial}{\partial t}+d_{31}\right)\tilde{\rho}_{31}+\frac12\zeta^*(z)\tilde{\rho}_{21}+\frac12\zeta(z)\Omega_\text{c}\tilde{\rho}_{41}-\frac12\zeta^*(z)\Omega_\text{p}\tilde{\rho}_{32}=&0,
    \end{align}
    \begin{align}
    \text{i}\left(\frac{\partial}{\partial t}+d_{41}\right)\tilde{\rho}_{41}+\frac12\zeta^*(z)\Omega_\text{c}\tilde{\rho}_{31}+\frac12\Omega_\text{m}^*\tilde{\rho}_{21}-\frac12\zeta^*(z)\Omega_\text{c}\tilde{\rho}_{42}=&0,
    \end{align}
    \begin{align}
    \text{i}\left(\frac{\partial}{\partial t}+d_{23}\right)\tilde{\rho}_{23}+\frac12\zeta^*(z)\Omega_\text{p}\tilde{\rho}_{13}-\frac12\zeta^*(z)\Omega_\text{c}\tilde{\rho}_{24}+\frac12\Omega_\text{m}\tilde{\rho}_{43}+\frac12\zeta(z)\Omega_\text{s}(\tilde{\rho}_{33}-\tilde{\rho}_{22})=&0,
    \end{align}
    \begin{align}
    \text{i}\left(\frac{\partial}{\partial t}+d_{42}\right)\tilde{\rho}_{42}+\frac12\zeta(z)\Omega_\text{p}\tilde{\rho}_{41}-\frac12\zeta^*(z)\Omega_\text{c}\tilde{\rho}_{32}+\frac12\zeta^*(z)\Omega_\text{s}\tilde{\rho}_{43}+\frac12\Omega_\text{m}^*(\tilde{\rho}_{22}-\tilde{\rho}_{44})=&0,
    \end{align}
    \begin{align}
    \text{i}\left(\frac{\partial}{\partial t}+d_{43}\right)\tilde{\rho}_{43}+\frac12\Omega_\text{m}^*\tilde{\rho}_{23}+\frac12\zeta(z)\Omega_\text{s}\tilde{\rho}_{42}+\frac12\zeta^*(z)\Omega_\text{c}(\tilde{\rho}_{33}-\tilde{\rho}_{44})=&0,
\end{align}
here 
\begin{align}
    d_{21}
        =&-\Delta\omega_{21}+\Delta_\text{p}+\text{i}\gamma_{21},\;
    d_{31}=-\Delta\omega_{31}+(\Delta_\text{p}-\Delta_\text{s})+\text{i}\gamma_{31},\; d_{41}=-\Delta\omega_{41}+(\Delta_\text{c}+\Delta_\text{p}-\Delta_\text{s})+\text{i}\gamma_{41},\\
    d_{23}=&-\Delta\omega_{23}+\Delta_\text{s}+\text{i}\gamma_{23},\;d_{42}=-\Delta\omega_{42}+(\Delta_\text{c}-\Delta_\text{s})+\text{i}\Gamma_{42},\; d_{43}=-\Delta\omega_{43}+\Delta_{c}+\text{i}\gamma_{43},
\end{align}
and 
\begin{align}
    \Gamma_{33}:=\Gamma_{23}-\Gamma_{31},\; \Gamma_{44}:=\Gamma_{43}+\Gamma_{42}+\Gamma_{41},
\end{align}
with $\Gamma_j:=\sum_{l\neq j}\Gamma_{jl}$ the total decay rate of the $\ket{j}$ level, $\gamma_{ij}^{\text{dep}}$ as dephasing rates of the correspond atomic transitions and
\begin{equation}
   \gamma_{ij}=\frac{\Gamma_{i}+\Gamma_j}{2}+\gamma_{ij}^{\text{dep}}.
\end{equation}
The Liouville equation for left and right probe field polarizations can then be readily obtained by mapping
\begin{align}
    \Omega_\text{p}\mapsto\Omega_\text{p}^{\pm},\;\;\;\Delta_\text{p}\mapsto\Delta_\text{p}^{\pm}.
\end{align}

The inhomogeneous broadening of the solid sample affect $d_{ij}$ due to small-energy shift $\Delta\omega_{ij}$ of the $\ket{i}\leftrightarrow\ket{j}$ atomic transition. Investigating the effect of the inhomogeneous broadening to the GVD, SPM, XPM and other nonlinear parameters of the system are challenging and need further considerations. Our predicted phase singularities and nonlinear dark rogue waves could be achieved in a specific subensemble of the $\text{Pr}^{3+}$-ions with $\Delta\omega_{ij}\mapsto0$ such as those occur via persistent spectral-hole burning technique. 
\section{\label{Exact First order} Analytic solution of the first-order perturbative solution}
In this appendix, we deal with the first order perturbative solution of the Maxwell-Bloch equation. It is worth noting that the perturbative solution of the Maxwell-Bloch equations yields set of linear and inhomogeneous equations that can be solved order by order.  

 We obtain the first-order perturbative solution by linearizing the coupled Maxwell-Bloch equations. This linearization is achieved by substituting Eq.~(\ref{Eq:Linear_Density_Matrix}) and Eq.~(\ref{Eq:Linear_Rabi_Frequency}) into Maxwell-Bloch equations. Then, we obtain the first-order solution of the density-matrix elements in this hybrid waveguide can be expressed as
\begin{align}
    a_{21}^{\pm1}=&\frac{|\zeta(z)\Omega_\text{c}|^2+(d_{31}^{\pm}-\text{i}\omega)(d_{41}^{\pm}-\text{i}\omega))}{\Omega_\text{s}\Omega_\text{c}(\bar{\Omega_\text{m}}^2)-|\zeta(z)|^2[\Omega_\text{c}^2(\omega+\text{i}d_{21}^{\pm})+\Omega_\text{s}^2(\omega+\text{i}d_{41}^{\pm})]+(\omega+\text{i}d_{31}^{\pm})[(\omega+\text{i}d_{21}^{\pm})(\omega+\text{i}d_{41}^{\pm})-|\Omega_\text{m}|^2]},\label{Eq:Linear_Dispersion_Coefficient}\\
    a_{31}^{\pm1}=&\frac{\zeta^*(z)\Omega_\text{s}(\text{i}d_{41}^{\pm}+\omega)-\zeta(z)\Omega_\text{c}\Omega_\text{m}^*}{\Omega_\text{s}\Omega_\text{c}(\bar{\Omega_\text{m}}^2)-|\zeta(z)|^2
    [\Omega_\text{c}^2(\omega+\text{i}d_{21}^{\pm})+\Omega_\text{s}^2(\omega+\text{i}d_{41}^{\pm})]+(\omega+\text{i}d_{31}^{\pm})[(\omega+\text{i}d_{21}^{\pm})(\omega+\text{i}d_{41}^{\pm})-|\Omega_\text{m}|^2]},\\
    a_{41}^{\pm1}=&\frac{\Omega_\text{m}^*(\text{i}d_{31}+\omega)-\zeta^{*2}(z)\Omega_\text{c}\Omega_\text{s}}{\Omega_\text{s}\Omega_\text{c}(\bar{\Omega_\text{m}}^2)-|\zeta(z)|^2[\Omega_\text{c}^2(\omega+\text{i}d_{21}^{\pm})+\Omega_\text{s}^2(\omega+\text{i}d_{41}^{\pm})]+(\omega+\text{i}d_{31}^{\pm})[(\omega+\text{i}d_{21}^{\pm})(\omega+\text{i}d_{41}^{\pm})-|\Omega_\text{m}|^2]}.
\end{align}
Above analytic solutions describe the optical properties of the hybrid-plasmonic waveguide in the linear approximation. The chromatic linear dispersion of this hybrid waveguide would then obtained using Eq.~(\ref{Eq:Linear_Dispersion_Coefficient}) for both left-right circular polarization of the probe field.  
\section{\label{Derivation of dark coupler rogue waves} Dressed darboux transformation and coupled dark rogue wave solution}
We obtain the Manakov system based on coupled nonlinear Schr\"odinger equation formalism (i.e., Eq.~(\ref{Eq:manakov})) by mapping
\begin{align}
    u^{\imath}\exp\left\{\text{i}\frac{g_\text{d}^\imath }{2g_\text{D}^\imath }\left(t-\frac{g_\text{d}^\imath }{4}x\right)\right\}\mapsto\tilde{u}^\imath ,
\end{align}
assuming normal GVD and considering symmetric DEIT windows. Then the Manakov system reduces to $3\times3$ linear eigenvalue problem 
\begin{align}
    \frac{\partial\bm{R}}{\partial \sigma}=(\Lambda\bm{G}+\bm{\mathcal{Q}})\bm{R},\; \frac{\partial\bm{R}}{\partial s}=\left[-\frac{3}{2}\Lambda^2\bm{G}-\frac{3}{2}\Lambda\bm{Q}+\frac{\text{i}}{2}\sigma_3\left(\bm{Q}^2-\frac{\partial\bm{Q}}{\partial \sigma}\right)\right]\bm{R},
    \label{Plasmonic_Lax-Pairs}
\end{align}
$\Lambda$ is the spectral parameter, $\bm{R}:=(r,s,t)^{\dagger}$ is the arbitrary matrix ($\dagger$ denotes the matrix transpose),
\begin{align}
    \bm{G}=\text{diag}(-2\text{i},\text{i},\text{i}),\; \sigma_3=\text{diag}(1,-1,-1)
    \label{Eq:Law Pair}
\end{align}
and
\begin{align}
 \bm{\mathcal{Q}}=\left(
    \begin{matrix}
     0&u^+&u^-\\
     u^{+*}&0&0\\
     u^{-*}&0&0\\
    \end{matrix}\right).
\end{align}
We assume the two-mode plane SPP waves excite and propagate as
\begin{equation}
    u_0^{\pm}(x,t)=|u_0^{\pm}|\exp\{\text{i}\Omega^{\pm}(\kappa^{\pm}\tilde{x}-\tau)\}
\end{equation}
with
\begin{align}
    \tau:=\tau_0(t-x/v_+),\; \tilde{x}:=x\left(k(\omega)+K(\omega)+\frac{1}{L_\text{D}}\right).
\end{align}
In order to obtain the general solution of the Lax-pairs, we introduce 
\begin{align}
    \mathcal{G}:=\text{diag}\left(1,\frac{u^{+*}}{|u_{0}^{+}|},\frac{u^{-*}}{|u_{0}^{-}|}\right),\;\;\bm{\Phi}:=\bm{\mathcal{G}}^{-1}\bm{R}(\Lambda),
\end{align}
Then we assume matrix perturbation around a stable pole $\Lambda\approx\Lambda_{0}$ as
\begin{align}
     \Lambda(\varepsilon)\approx&\Lambda_0+(\Lambda_0-\Lambda_0^*)\varepsilon^2,\\
    \bm{\Phi}(\Lambda)\approx&\bm{\Phi}^{(0)}+\varepsilon^2\bm{\Phi}^{(1)}+\varepsilon^{4}\bm{\Phi}^{(2)}+\cdots+\mathcal{O}(\varepsilon^{2n}).
\end{align}
We define the characteristic matrix $\bm{Y}_{1\times n}$ as
\begin{equation}
    \left[\begin{matrix}
    \bm{Y}_1\\
    \bm{Y}_2\\
    \bm{Y}_3
    \end{matrix}\right]=
    \left[\begin{matrix}
    \bm{\Phi}^{(0)},\bm{\Phi}^{(1)},\bm{\Phi}^{(2)},\ldots,\bm{\Phi}^{(n-1)}
    \end{matrix}\right],
\end{equation}
and evaluate the elements of the $\bm{M}$ matrix elements as
\begin{equation}
    \frac{\bm{\Phi}^{\dagger}\bm{\sigma}_3\bm{\Phi}}{\Lambda-\Lambda_{0}}=\sum_{i,j}^nM_{ij}\varepsilon^{*2(i-1)}\varepsilon^{2(j-1)}+\mathcal{O}(\varepsilon^{4n}).
\end{equation}
We then obtain the $n^\text{th}$ order solution of Eqs.~\ref{Plasmonic_Lax-Pairs} as
\begin{align}
    u^{+[n]}=u_{0}^{+}\left[1+\frac{3\text{i}}{|u_{0}^{+}|}\bm{Y}_1\bm{M}^{-1}\bm{Y}_2^{\dagger}\right],\\
    u^{-[n]}=u_{0}^{-}\left[1+\frac{3\text{i}}{|u_{0}^{-}|}\bm{Y}_1\bm{M}^{-1}\bm{Y}_3^{\dagger}\right].
\end{align}
Employing the Darboux transformation~\cite{PhysRevE.89.041201} in the weak perturbation of the spectral parameter yields to first-order coupled-dark rogue wave generation described by
\begin{align}
    u^\imath (s,\sigma)
        =u_0^{\imath}\left[1+\frac{3(\Lambda_0-\Lambda_0^*)\vartheta^{\imath*}\vartheta/\beta^{\imath*}}{|\vartheta|^2-(|u_0^+||\vartheta^+/\beta^+|^2+|u_0^-||\vartheta^-/\beta^-|^2})\right],
    \label{Eq:Dark Rogue wave}
\end{align}
with
\begin{align}
    \eta^{\imath2}=4\sqrt{|u_0^\imath |^2(|u_0^\imath |^2+\delta^2)}-(4|u_0^\imath |^2-\delta^2),\;\;\delta=\omega_*^+-\omega_*^-,\;\;\omega_*=\omega_*^+-\omega_*^+.
\end{align}
here, we also assume $\eta^+\approx\eta^-:=\eta$ and
\begin{align}
    \vartheta:=\sigma-(\omega_*+\text{i}\eta)s/2,\;\beta^\imath :=(\omega_*+\text{i}\eta)/2-\omega_*^\imath , \vartheta^\imath :=\vartheta+\frac{1}{\text{i}\beta^\imath }.
\end{align}
Our simulations are performed by rotating Eq.~(\ref{Eq:Dark Rogue wave}) to the original variable $(x,\tau)$.
\section{\label{DerivationCoupled}Derivation of the coupled nonlinear Schr\"{o}dinger equation}
In this section, we derive the coupled-mode theory for SPP waves at the nonlinear medium-metamaterial layer at the interface.
We start with Maxwell equations
\begin{align}
    \bm{\nabla}\times\bm{E}
=-\frac{\partial \bm{B}}{\partial t},\quad \bm{\nabla}\times \bm{H}=\bm{J}+\frac{\partial \bm{D}}{\partial t}.
\end{align}
We assume no charge density
(i.e., $\bm{J}=0$), monochromatic electromagnetic fields with angular frequency ($\omega$).
Treating optical properties of nonlinear medium as $\tilde{\varepsilon}$,
$\mu$, electromagnetic fields with $\tilde{\bm{E}}$, $\tilde{\bm{H}}$, 
\begin{align}
    \tilde{\bm{B}}=\mu\tilde{\bm{H}},\quad\tilde{\bm{D}}=\tilde{\varepsilon}\tilde{\bm{E}},
\end{align}
substituting into Maxwell equations we achieve
\begin{align}
    \bm{\nabla}\times\bm{E}=\text{i}\omega\mu\bm{H},\quad \bm{\nabla}\times\bm{H}=-\text{i}\omega\varepsilon\bm{E},\\
    \bm{\nabla}\times\tilde{\bm{E}}=\text{i}\omega\mu\tilde{\bm{H}},\quad \bm{\nabla}\times\tilde{\bm{H}}=-\text{i}\omega\tilde{\varepsilon}\tilde{\bm{E}}.
\end{align}
Defining
\begin{equation}
    \bm{\mathcal{F}}:=\bm{E}\times\tilde{\bm{H}}^*+\tilde{\bm{E}}^*\times\bm{H},
\end{equation}
and using the divergent formula as
\begin{equation}
    \bm{\nabla}\cdot\left(\bm{A}\times\bm{B}\right)=\bm{\nabla}\cdot\bm{A}\times\bm{B}+\bm{A}.\bm{\nabla}\times\bm{B},
\end{equation}
we obtain the divergent in $\bm{\mathcal{F}}$ (i.e., $\bm{\nabla}\cdot\bm{\mathcal{F}}$) as
\begin{equation}
    \bm{\nabla}\cdot\bm{\mathcal{F}}=\text{i}\omega\left(\varepsilon-\tilde{\varepsilon}^*\right)\bm{E}\cdot\bm{\tilde{E}}^*.
\end{equation}
We also use the divergence theorem to a small volume with infinitesimal thickness $z$ and integration area $\mathcal{A}$,
\begin{equation}
    \int_\mathcal{A}\bm{\nabla}\cdot\bm{\mathcal{F}}\,\text{d}\mathcal{A}=\frac{\partial}{\partial z}\int_\mathcal{A}\bm{\mathcal{F}}\cdot\hat{n}\,\text{d}\mathcal{A}+\oint_{l_\text{A}}\bm{\mathcal{F}}\cdot\hat{n}\,\text{d}l_\mathcal{A},
    \label{identity}
\end{equation}
consider the electromagnetic modes of the metallic layer to be expand as
\begin{align}
    \bm{E}_\text{p}(\bm{r}):=\bm{e}_\text{p}(x,y)\exp\{\text{i}\beta_\text{p}z\},\quad \bm{H}_\text{p}(\bm{r}):=\bm{h}_\text{p}(x,y)\exp\{\text{i}\beta_\text{p}z\}
\end{align}
where $p$ is the number of mode, $\bm{e}_\text{p}$ ($\bm{h}_\text{p}$) are the vector functions of the electrical~(magnetic) modes. Moreover, assuming
\begin{align}
    \beta_{-p}=-\beta_\text{p},\quad\bm{h}_{-p}=-\bm{h}_\text{p},\quad\bm{e}_{-p}=\bm{e}_\text{p}
\end{align}
and taking into the account the orthogonality of the electromagnetic modes as
\begin{align}
    \int_{\mathcal{A}_{\infty}}\hat{z}\cdot\bm{e}_\text{p}\times\bm{h}_{q}^*\,\text{d}\mathcal{A}=\pm\delta_{pq},
\end{align}
we rewrite the deterministic function ($\mathcal{F}$) as
\begin{align}
    \bm{\mathcal{F}}_\text{p}=\bm{e}_\text{p}\times\tilde{\bm{H}}^*+\tilde{\bm{E}}\times\bm{h}_\text{p}^*.
\end{align}
Substituting this result into Eq.~(\ref{identity}) we achieve
\begin{equation}
    \left(\text{i}\omega\int(\varepsilon-\tilde{\varepsilon}^*)\bm{e}_\text{p}\cdot\tilde{\bm{E}}^*\,\text{d}\mathcal{A}-\oint_{l_\mathcal{A}}\bm{\mathcal{F}}\cdot\hat{n}\,\text{d}l_\mathcal{A}\right)\exp\{\text{i}\beta_\text{p}z\}=\frac{\partial}{\partial z}\int_\mathcal{A}\exp\{\text{i}\beta_\text{p}z\}\left(\bm{e}_\text{p}\times\tilde{\bm{H}}^*+\tilde{\bm{E}}\times\bm{h}_\text{p}^*\right)\cdot\hat{n}\,\text{d}\mathcal{A},
\end{equation}
simplifying the result yields
\begin{equation}
    \left(\frac{\partial}{\partial z}+\text{i}\beta_\text{p}\right)\int_\mathcal{A}\bm{\mathcal{F}}\cdot\hat{z}\,\text{d}\mathcal{A}=\text{i}\omega\int(\varepsilon-\tilde{\varepsilon}^*)\bm{e}_\text{p}\cdot\tilde{\bm{E}}^*\,\text{d}\mathcal{A}-\oint_{l_\mathcal{A}}\bm{\mathcal{F}}\cdot\hat{n}\,\text{d}l_\mathcal{A}.
    \label{simplified}
\end{equation}
In the hybrid plasmonic waveguide, the electromagnetic field \text{exponentially decay} at large distance of the interface, i.e.
\begin{equation}
    \lim_{l_\mathcal{A}\rightarrow\infty}\oint_{l_\mathcal{A}}\bm{\mathcal{F}}\cdot\hat{n}\,\text{d}l_\mathcal{A}\mapsto0.
    \label{intapprox}
    \end{equation}
Moreover, we consider the electromagnetic fields in the nonlinear layer to be expanded in terms of normal modes $\bm{e}_{q\text{t}}$ ($\bm{h}_{q\text{t}}$) and slowly varying amplitudes ($a_\text{p}$) as
\begin{align}
    \tilde{\bm{E}}=\sum_{q}a_{q}(z)\bm{e}_{q\text{t}},\quad \tilde{\bm{H}}=\sum_{q}a_{q}(z)\bm{h}_{q\text{t}}.
    \label{modenonlinearexpandion}
\end{align}
Using Eqs.~(\ref{intapprox}) and~(\ref{modenonlinearexpandion}) in Eq.~(\ref{simplified}) results in basic equation of coupled-mode theory
\begin{equation}
    \frac{\text{d}a_\text{p}(z)}{\text{d}z}+\text{i}\beta_\text{p}a_\text{p}(z)=\frac{\text{i}\omega}{2}\int(\varepsilon-\tilde{\varepsilon}^*)\bm{e}_\text{p}\cdot\tilde{\bm{E}}^*\,\text{d}\mathcal{A}
\end{equation}
This equation can readily be adapted to the Mankov system using nonlinear polarization and employing two SPP modes. To this aim, we assume 
\begin{align}
    \text{i}\omega\int(\varepsilon-\tilde{\varepsilon}^*)\bm{e}_\text{p}\cdot\tilde{\bm{E}}^*\,\text{d}\mathcal{A}=-\text{i}\omega\int\bm{e}_\text{p}\cdot\tilde{\bm{P}}_\text{NL}\,\text{d}V,
\end{align}
consider the two-mode case $p\in\{1,2\}$, take into account the effect of SPM and XPM as
\begin{align}
    \tilde{\bm{P}}_\text{NL}
    	:=\chi_\text{11}^{(3)}|\bm{E}_1\cdot\bm{E}_1^*|\bm{E}_1+\frac{\chi_\text{12}^{(3)}}{2}|\bm{E}_2\cdot\bm{E}_2^*|\bm{E}_1,
\end{align}
here, $\chi_{12}^{(3)}\approx\chi_{21}^{(3)}$, $\chi_{11}^{(3)}\chi_{22}^{(3)}=\chi_{12}^{(3)}\chi_{21}^{(3)}$ and
\begin{align}
    W_\text{SPM}=&-\frac{1}{d}\iint\,\text{d}x\text{d}y\int\bm{e}_1\cdot\left[\chi_\text{11}^{(3)}|\bm{e}_1\cdot\bm{e}_1^*|\bm{e}_1+\frac{\chi_\text{12}^{(3)}}{2}|\bm{e}_1\cdot\bm{e}_1^*|\bm{e}_1^*\right]\,\text{d}z,\\
    W_\text{XPM}=&\frac{1}{d}\iint\,\text{d}x\text{d}y\int\bm{e}_1\cdot\left[\chi_\text{11}^{(3)}|\bm{e}_1\cdot\bm{e}_2^*|\bm{e}_1+\chi_\text{22}^{(3)}|\bm{e}_2\cdot\bm{e}_1^*|\bm{e}_2+\chi_\text{21}^{(3)}|\bm{e}_2\cdot\bm{e}_1^*|\bm{e}_2^*\right]\,\text{d}z,
\end{align}
which then leads to the plasmonic version of the Manakov system
\begin{align}
    \frac{\text{d}a_1(z)}{\text{d}z}-\text{i}\beta_1(\omega)a_1(z)=&\text{i}\left(W_\text{SPM}\left|a_1\right|^2+W_\text{XPM}\left|a_2\right|^2\right)a_1(z),\\
    \frac{\text{d}a_2(z)}{\text{d}z}-\text{i}\beta_2(\omega)a_2(z)=&\text{i}\left(W_\text{SPM}\left|a_2\right|^2+W_\text{XPM}\left|a_1\right|^2\right)a_2(z).
\end{align}
\end{widetext}
\end{document}